\documentclass[twocolumn,prb,aps,superscriptaddress]{revtex4-1}
\usepackage[T1]{fontenc}
\usepackage[latin9]{inputenc}

\usepackage{amsmath}
\usepackage{amssymb}
\usepackage{bbm}
\usepackage{braket}
\allowdisplaybreaks

\usepackage{graphicx}
\usepackage[colorlinks=true,citecolor=blue,linkcolor=blue]{hyperref}

\newcommand{\equref}[1]{Eq.~(\ref{#1})}
\newcommand{\equsref}[2]{Eqs.~(\ref{#1}) and (\ref{#2})}
\newcommand{\secref}[1]{Sec.~\ref{#1}}
\newcommand{\figref}[1]{Fig.~\ref{#1}}
\newcommand{\refcite}[1]{Ref.~\onlinecite{#1}}
\newcommand{\refscite}[1]{Refs.~\onlinecite{#1}}
\newcommand{\refsacite}[2]{Refs.~\onlinecite{#1} and \onlinecite{#2}}
\newcommand{\tableref}[1]{Table~\ref{#1}}
\newcommand{\supl}{Supplementary Information}
\newcommand{\diff}{\mathrm{d}}
\newcommand{\der}[2]{\frac{\mathrm{d} #1}{\mathrm{d} #2}}

\newcommand{\sign}{\,\text{sign}}
\renewcommand{\approx}{\simeq}

\renewcommand{\vec}[1]{\boldsymbol{#1}}

\begin{document}
\title{Topological superconductivity and unconventional pairing in oxide interfaces}
\author{Mathias S. Scheurer}
\affiliation{Institut für Theorie der Kondensierten Materie, Karlsruher Institut für Technologie, Karlsruhe, D-76131, Germany}

\author{Jörg Schmalian}
\affiliation{Institut für Theorie der Kondensierten Materie, Karlsruher Institut für Technologie, Karlsruhe, D-76131, Germany}
\affiliation{Institut für Festköperphysik, Karlsruher Institut für Technologie, Karlsruhe, D-76131, Germany}
\date{\today}

\begin{abstract}
To pinpoint the microscopic mechanism for superconductivity has proven
to be one of the most outstanding challenges in the physics of correlated
quantum matter. Thus far, the most direct evidence for an electronic
pairing mechanism is the observation of a new symmetry of the order-parameter,
as done in the cuprate high-temperature superconductors. Like distinctions
based on the symmetry of a locally defined order-parameter, global,
topological invariants allow for a sharp discrimination between states
of matter that cannot be transformed into each other adiabatically.
Here we propose an unconventional pairing state for the electron fluid
in two-dimensional oxide interfaces and establish a direct link to the emergence of nontrivial topological invariants.
Topological superconductivity and Majorana edge states can then be
used to detect the microscopic origin for superconductivity. In addition,
we show that also the density wave states that compete with superconductivity
sensitively depend on the nature of the pairing interaction. Our conclusion
is based on the special role played by the spin-orbit coupling
and the shape of the Fermi surface in SrTiO$_{3}$/LaAlO$_{3}$-interfaces
and closely related systems. 
\end{abstract}
\maketitle

The two-dimensional electron fluid that forms\cite{Ohtomo2004} at
the interface between the insulators SrTiO$_{3}$ and LaAlO$_{3}$
is an example of an engineered quantum system, where a new state
of matter emerges as one combines the appropriate building blocks.
The subsequent discovery of superconductivity\cite{Reyren2007} in
the interface, along with the ability to control the ground state
via applied electric fields\cite{Caviglia2008} opened up intense
research. The key open question is whether electronic correlations
promote new states of matter, such as unconventional superconductivity
or novel magnetic states\cite{Brinkman2007,Li2011,Bert2011,Banerjee2013,Li2014}
and how such phases are related to each other.

New states of matter can be sharply distinguished from conventional
behavior when they break a symmetry or differ in their topology. The
nontrivial consequences of the mapping from momentum space to the
space of Hamiltonians, as found in topological insulators and superconductors,
have recently had a major impact on solid state physics\cite{HasanKane,Qi}.
Here we propose a new electronic pairing mechanism for superconductivity
in oxide interfaces that is due to the exchange of particle-hole excitations
and that leads to topological superconductivity with Majorana bound
states and related nontrivial topological aspects. Specifically, we
find a time-reversal preserving topological superconductor that has
attracted recent attention\cite{ZhangKane,Keselman,Nakosai,Nakosai2,Deng,Fu}.
In contrast, conventional electron-phonon coupling in the same system
would lead to a topologically trivial state. We also study competing
states, expected to emerge nearby superconductivity in the phase
diagram. For a conventional pairing mechanism we find charge density
wave order, while an in-plane spin density wave with magnetic vortices
competes with unconventional superconductivity.

\begin{figure*}[ht]
\begin{center}
\includegraphics[width=0.75\linewidth]{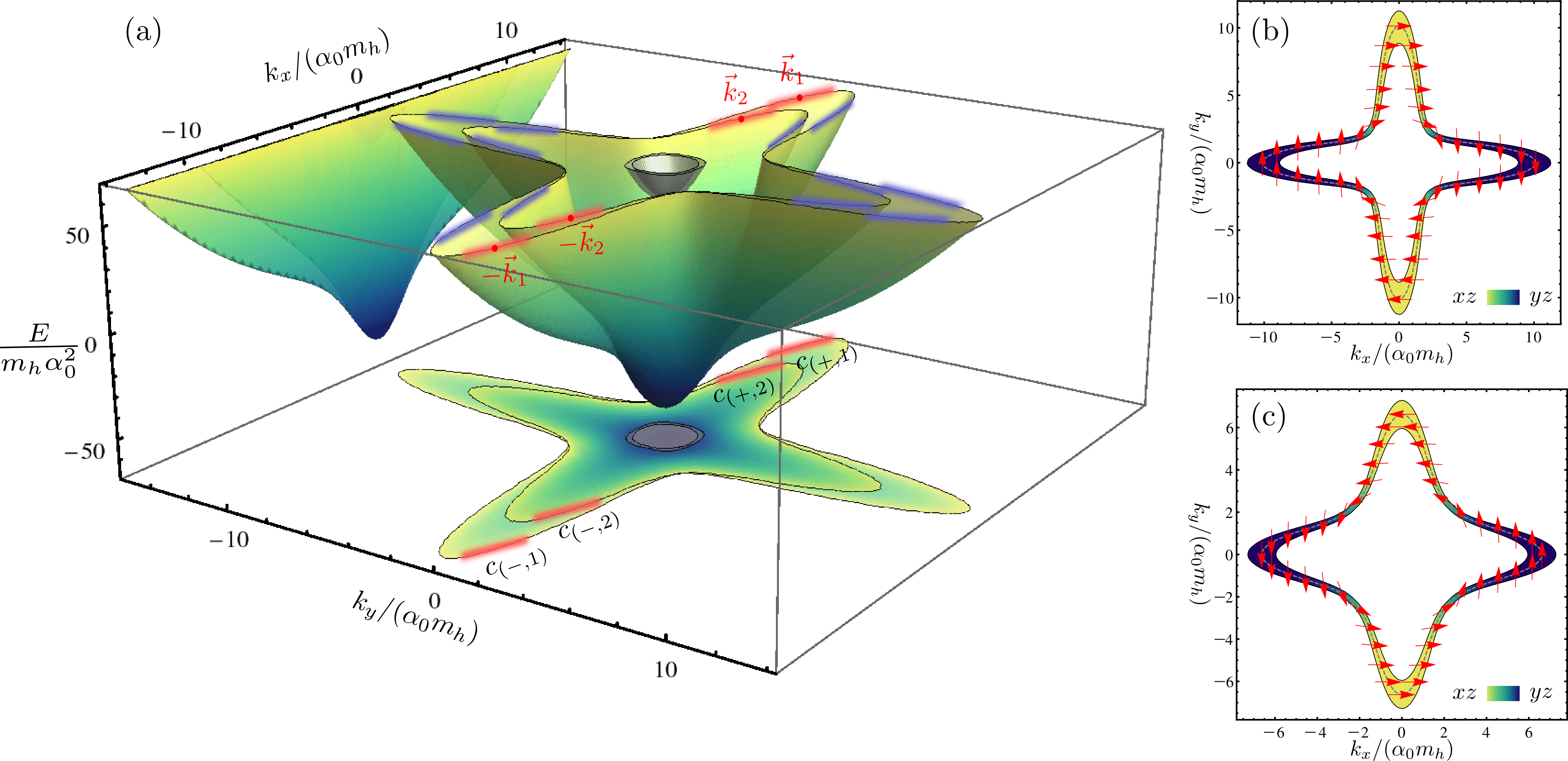}
\caption{Part (a) shows the spectrum of the effective two-band Hamiltonian using the realistic parameter stated in the main text. In this paper, we restrict the analysis on the 4 most strongly nested subspaces (highlighted in red and blue). The orbital weight (color) and orientation of the spin (red arrow) are illustrated in (b) and (c) for the outer and inner Fermi surface, respectively. Note that, as a consequence of time-reversal and $\pi$-rotation symmetry about the $z$-axis, the spin has to lie in the $xy$-plane.}
\label{SpectrumAndWavefunctions}
\end{center}
\end{figure*}

\section{Interacting low-energy model}
The crucial states near the Fermi energy of the oxide interface are
made up of titanium $3d_{xz}$ and $3d_{yz}$ orbitals\cite{Santander-Syro2011,Joshua2012,King2014}.
The orientation of the electron clouds of the $3d$-orbitals leads
to a wave function overlap along the $x$-direction that is much larger
for $d_{xz}$ states compared to $d_{yz}$, and vice versa for the
$y$-direction. Each orbital is then characterized by a light mass
$m_{l}$ and a heavy mass $m_{h}$, leading to the experimentally
observed strongly anisotropic electronic structure\cite{Santander-Syro2011,King2014}.
For example, the energy of the $d_{xz}$ states can be described by
\begin{equation}
\varepsilon_{xz}\left(k\right)=\frac{k_{x}^{2}}{2m_{l}}+\frac{k_{y}^{2}}{2m_{h}}\label{eq:disp}
\end{equation}
where $m_{h}/m_{l}\approx15\cdots30$. $\varepsilon_{yz}$ follows
from \equref{eq:disp} by interchanging $k_{x}$ and $k_{y}$. In
addition, the electronic properties of the polar interface between
insulating oxides is strongly affected by the spin-orbit interaction.
Due to the Dresselhaus-Rashba effect\cite{Dresselhaus1955,Rashba1960},
the electronic states experience a momentum dependent splitting and
mixing of spin-states, naturally explaining magneto-transport experiments\cite{Caviglia2010,BenShalom2010}.
The effect might also be responsible for the observed phase separation
in interfaces\cite{Bucheli2013}. Focusing on the $d_{xz}$ and $d_{yz}$
states, the most general form up to linear order in momentum that
is consistent with the $C_{4v}$-point group symmetry and time-reversal
invariance is given by
\begin{align}
\begin{split}H_{\text{so}}(\vec{k}) & =\frac{1}{2}\lambda\tau_{2}\sigma_{3}+\alpha_{0}\tau_{0}\left(k_{x}\sigma_{2}-k_{y}\sigma_{1}\right)\\
 & +\alpha_{1}\tau_{1}\left(k_{x}\sigma_{1}-k_{y}\sigma_{2}\right)+\alpha_{3}\tau_{3}\left(k_{x}\sigma_{2}+k_{y}\sigma_{1}\right),\label{eq:spinorbfin}
\end{split}
\end{align}
where the Pauli matrices $\sigma_{i}$ and $\tau_{j}$ ($i,j=0,\dots3$)
act in spin and orbital space, respectively. Projecting out the $d_{xy}$
band that is closest in energy and shifted by $\epsilon_{0}$ and
including the atomic spin-orbit coupling $H_{\text{so}}=\lambda\vec{L}\cdot\vec{s}$
we find $\alpha_{0}=-\alpha_{1}=-\alpha_{3}=\frac{1}{2}\delta\lambda/\epsilon_{0}$.
$\lambda\approx20\:{\rm meV}$, $\delta/a_{0}\approx40\,{\rm meV}$,
and $\epsilon_{0}\approx250\,{\rm meV}$ were determined in first
principles calculations\cite{Zhong2013}. As $\delta$ and $\epsilon_{0}$
depend sensitively on details of the interface we use $\alpha_{0}\approx10\cdots50\,{\rm meV}\mathring{{A}}$,
estimated from magnetotransport experiments\cite{Caviglia2010}. 

In \figref{SpectrumAndWavefunctions}(a) we show the bands that result
from the combination of the anisotropic masses in \equref{eq:disp} and the
spin-orbit coupling (\ref{eq:spinorbfin}). Two of the four bands
are pushed to higher energies by the atomic spin-orbit coupling $\lambda\tau_{2}\sigma_{3}$
and can thus be neglected for the following low-energy analysis as
long as the chemical potential is tuned sufficiently far away from
the bottom of these bands. The remaining two bands are split by the
Dresselhaus-Rashba coupling and show strong nesting in the highlighted
regions. We emphasize the similarity of the Fermi surface to the one
reported in \refcite{King2014} for the surface states of SrTiO$_{3}$.
The nesting is a consequence of the mass anisotropy and becomes exact
in the limit $m_{l}/m_{h}\rightarrow0$.

This allows us to use a low-energy theory that involves only the degrees
of freedom in the vicinity of the most parallel slices of the Fermi
surface. In total, there are four equivalent strongly nested subspaces
that are related by the point group symmetries. Without loss of generality,
let us focus on, e.g., the one indicated in red in \figref{SpectrumAndWavefunctions}(a).
In this subset of momentum space, we introduce helicity creation and
annihilation operators $c_{(\sigma,j)}$ and $c_{(\sigma,j)}^{\dagger}$
which diagonalize the quadratic part of the Hamiltonian. Here $\sigma=\pm$
refers to the sign of $k_{x}$ and $j=1$ ($j=2$) denotes the outer
(inner) Fermi surface. To relate these operators to observables, \figref{SpectrumAndWavefunctions}(b)
and (c) show the spin-orientation and the orbital weight of the states
in the vicinity of the outer and inner Fermi surface, respectively.

There are two types of interaction processes allowed by momentum conservation
which we will refer to as backscattering and forward scattering. The
most general momentum independent backscattering term is given by
\begin{equation}
H_{\text{back}}=\sum_{s,s'=0}^{3}\sum_{\vec{q}}J_{s}^{-}(\vec{q})u_{s,s'}J_{s'}^{+}(-\vec{q}),\label{eq:backscattering}
\end{equation}
where ($\sigma=\pm$) 
\begin{equation}
J_{s}^{\sigma}(\vec{q})=\sum_{\vec{k},j,j'}c_{(\sigma,j)}^{\dagger}(\vec{k}+\vec{q})\left(\sigma_{s}\right)_{j,j'}c_{(\sigma,j')}(\vec{k}).\label{eq:currentdensity}
\end{equation}
We emphasize that from now on the Pauli matrices $\sigma_{s}$, as in \equref{eq:currentdensity},
do not describe the physical spin but rather act in the abstract isospin
space of the local helicity operators. The momentum of the operator
$c_{\sigma,j}$ is measured relative to the center $\sigma\vec{k}_{j}$
of the corresponding red region in \figref{SpectrumAndWavefunctions}(a).
Using the phase convention for the eigenstates defined in the \supl,
one finds that the $\pi$-rotation symmetry with respect to the $z$-axis
implies that $u$ has to be symmetric, $u^{T}=u$. The remaining symmetries
of the point group then fully determine the interaction in the other
three most strongly nested subspaces. In addition, time-reversal symmetry
imposes the constraint $u_{s,s'}=0$ if either $s=2$ or $s'=2$.
Let us first assume that the cutoff $\Lambda_{\perp}$ for the momenta
perpendicular to the Fermi surface can be chosen smaller than the
distance between the inner and outer Fermi surface. This means that
the red regions in \figref{SpectrumAndWavefunctions}(a) do not overlap.
In this situation, momentum conservation rules out further interaction
processes such that only $u_{00}$, $u_{11}=-u_{22}$, $u_{33}$ and
$u_{30}=u_{03}$ can be non-zero.

In case of forward scattering, where all four fermions have the same
index $\sigma$, the combination of Fermi statistics and point symmetries
leads to only one independent coupling constant.
 
\begin{figure}[tb]
\begin{center}
\includegraphics[width=\linewidth]{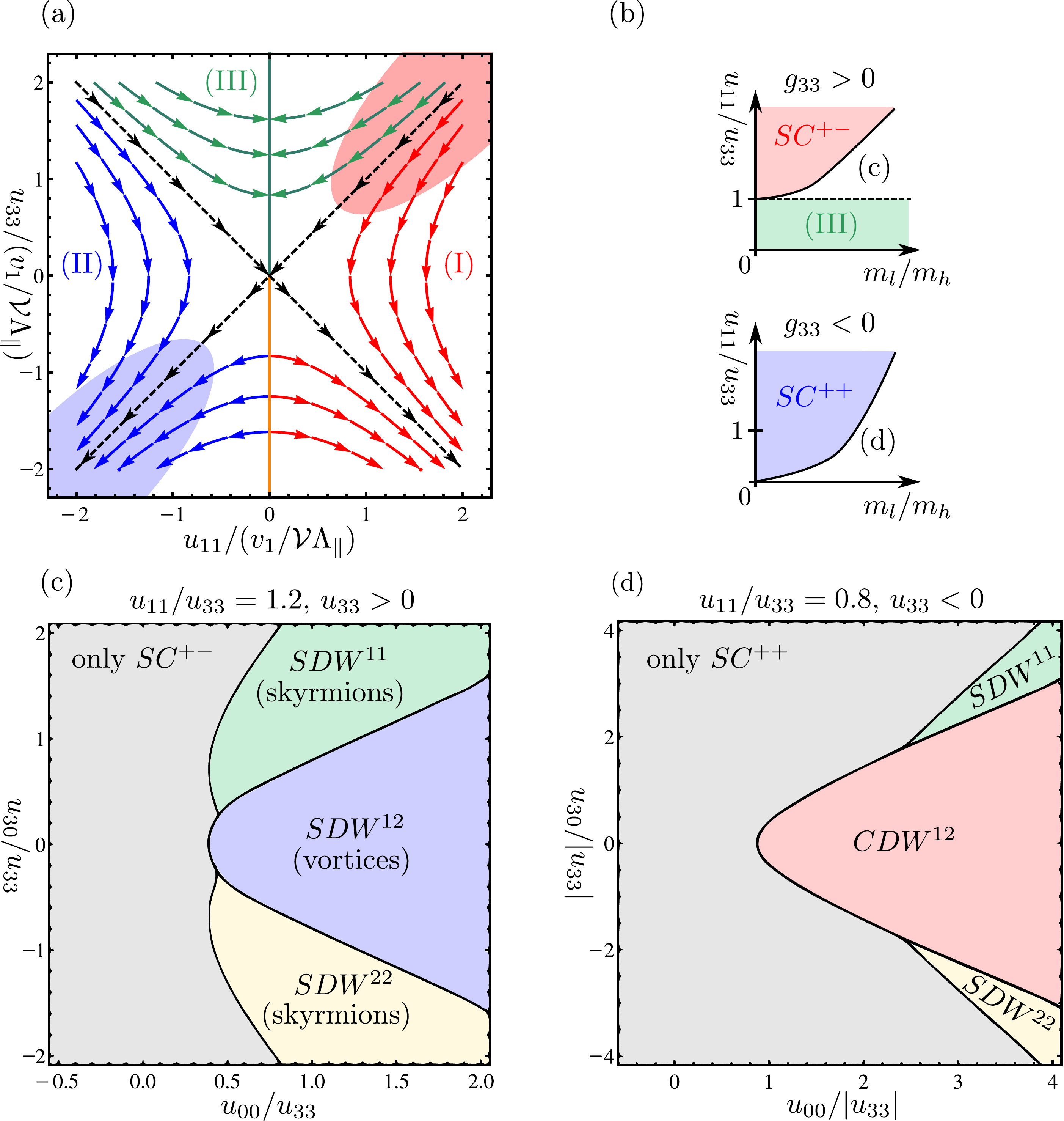}
\caption{In (a) the flow of the two running coupling constants in the case
of identical Fermi velocities is shown. Here $\mathcal{V}$ and $\Lambda_{\parallel}$
denote the volume of the system and the cutoff of the nested subspaces
tangential to the Fermi surface. Only in the regimes (I) and (II)
the couplings diverge indicating that the system develops an instability.
The red (blue) regions correspond to the bare couplings for a microscopically
repulsive (attractive) interaction. The schematic phase diagrams taking
into account finite mass anisotropies are shown in part (b). The non-flowing
couplings determine the properties of the charge density wave as shown
in (c) and (d) for the unconventional and conventional superconductor,
respectively. The nomenclature of the phases is explained in the main
text.}
\label{RGflowAndPhaseDiagram}
\end{center}
\end{figure}

\section{Pairing Instability and Topological Superconductivity}
Having derived the interacting low-energy Hamiltonian, we can now
deduce the associated instabilities. We perform a standard fermionic
one-loop Wilson renormalization group (RG) calculation\cite{Shankar}, in which high-energy
degrees of freedom are successively integrated out yielding an effective
Hamiltonian with renormalized coupling constants. If, during this
procedure, some of the couplings diverge, the system will develop
an instability. Following \refscite{Chubukov,Vafek} we identify
the physical nature of this instability by determining the order parameter
that has the highest transition temperature, allowing for all possible
(momentum independent) particle-hole and particle-particle ordered
states:
\begin{subequations}\begin{align}
\Delta_{\alpha,\beta}^{\text{DW}} & :=\sum_{\vec{k}}\braket{c_{\alpha}^{\dagger}(\vec{k})c_{\beta}(\vec{k})},\\
\overline{\Delta}_{\alpha,\beta}^{\text{SC}} & :=\sum_{\vec{k}}\braket{c_{\alpha}^{\dagger}(\vec{k})c_{\beta}^{\dagger}(-\vec{k})},
\end{align}\end{subequations}
where $\alpha$ and $\beta$ are double indices comprising helicity $\sigma=\pm$ and the Fermi surface sheet index $j=1,2$.
Near the Fermi surface, we linearize the band dispersion $\epsilon(\vec{k})\approx\pm v_{j}k_{\perp}$
with $k_{\perp}$ denoting the component of the momentum perpendicular
to the Fermi surface. For simplicity, let us first focus on the situation
$v_{1}=v_{2}$ which is quantitatively a good approximation even when
the chemical potential gets closer to the bottom of these bands. Below,
we will also discuss the more general case $v_{1}\neq v_{2}$. 

If $v_{1}=v_{2}$, only $u_{11}$ and $u_{33}$ out of the five coupling
constants flow as shown in \figref{RGflowAndPhaseDiagram}(a). We
find two regimes, denoted by (I) and (II), where the running couplings
diverge. In both cases, the instability is of superconducting type
characterized by the two non-zero anomalous expectation values $\Delta_{(-,j),(+,j)}^{\text{SC}}$
with $j=1,2$. As expected, we only have intra-Fermi surface pairing,
i.e.~only Kramer partners are paired. In region (I), the superconducting
order parameters of the nearby Fermi surfaces have opposite sign whereas
in (II) the sign is the same. The corresponding superconducting states
will be denoted by $SC^{+-}$ and $SC^{++}$, respectively. In the
region (III), none of the coupling constants diverge which means that,
for sufficiently small bare couplings, the system will not develop
any instability and, thus, reside in the metallic phase.

To unveil the microscopic pairing mechanism of the two superconducting
states, we start from a repulsive Coulomb interaction between the $d$-orbitals and
project onto the effective low-energy theory. This places us into
region (I) of the RG flow in \figref{RGflowAndPhaseDiagram}(a). In contrast, an attractive
interaction due to electron-phonon coupling would lead to initial couplings
in region (II). Consequently,
$SC^{++}$ results from conventional electron-phonon pairing, whereas
$SC^{+-}$ is an unconventional superconductor, where particle-hole
fluctuations effectively change the sign of $u_{33}$.

Both $SC^{+-}$ and $SC^{++}$ respect time-reversal symmetry as far
as the degrees of freedom of the nested subspaces are concerned. It
is natural to assume that this holds for the entire Fermi surface
and that, in addition, the system does not break the point symmetries relating the
nested segments. In this case the gap is finite
on the entire Fermi surface as seen in recent experiments\cite{Richter}.
Being fully gapped, it is natural to ask whether the time-reversal
invariant two-dimensional superconductor (class\cite{AltlandZirnbauer}
DIII) is topologically trivial or nontrivial\cite{Schnyder},
which is of great interest as it strongly influences its physical
properties. The most prominent feature of a nontrivial topological
superconductor is the appearance of spin-filtered counter propagating
Majorana modes at its edge when surrounded by a trivial phase\cite{BernevigsBook}.
It has been shown\cite{Zhang} that the associated topological invariant
$N\in\mathbbm{Z}_{2}$ is fully determined by the sign of the paring
field on the Fermi surfaces. It holds 
\begin{align}
N=\prod_{j}\left(\sign(\delta_{j})\right)^{m_{j}},\quad\delta_{j}=\braket{\psi_{j}|T\Delta_{j}^{\dagger}|\psi_{j}},\label{eq:TopInvar}
\end{align}
where the product involves all Fermi surfaces, $\psi_{j}$ and $\Delta_{j}$
denote the wave function of the non-interacting part of the Hamiltonian
and the pairing field at an arbitrary point on the $j$th Fermi surface.
Furthermore, $m_{j}$ is the number of time-reversal invariant points
enclosed by the $j$th Fermi surface and $T$ is the unitary part
of the time-reversal operator, given by $T=i\tau_{0}\sigma_{y}$ in
the basis of \equref{eq:spinorbfin}. As, in the present case, both
Fermi surfaces enclose only one time-reversal invariant point, the
superconductor is topological (trivial) if the sign of $\delta_{j}$
is different (identical) on the two Fermi surfaces. Inserting the order parameters derived above, we obtain
the pairing Hamiltonian
\begin{align}
H_{\text{pair}}=\Delta\sum_{\vec{k},j,j'} & c_{(+,j)}(\vec{k})\left(\gamma_{0}\sigma_{0}+\gamma_{3}\sigma_{3}\right)_{j,j'}c_{(-,j')}(-\vec{k})\nonumber \\
 & +\text{H.c.}
\end{align}
with $\gamma_{0}=u_{00}+2u_{11}+u_{33}$, $\gamma_{3}=2u_{30}$ for
the superconductor $SC^{++}$ and $\gamma_{0}=2u_{30}$, $\gamma_{3}=u_{00}-2u_{11}+u_{33}$
for the $SC^{+-}$-state. Calculating $\delta_{j}$ in \equref{eq:TopInvar},
one finds (see \supl~for details) that the superconductor is topological
if $|\gamma_{0}|<|\gamma_{3}|$ and trivial for the reversed inequality
sign. At $|\gamma_{0}|=|\gamma_{3}|$, the gap closes as is characteristic
for a topological phase transition. Recalling the flow depicted in
\figref{RGflowAndPhaseDiagram}(a), one immediately sees that $SC^{++}$
is trivial, whereas $SC^{+-}$ is a topological superconductor. Accordingly,
the experimental observation of topological features of the superconducting
state implies that the pairing mechanism must be unconventional as
it is the case for $SC^{+-}$. Vice versa, a trivial state is only
consistent with conventional, electron-phonon induced superconductivity.

We emphasize the difference of this result to recent work\cite{Mohanta,Fidkowski1,Fidkowski2,Kim}
proposing the emergence of Majorana fermions in the heterostructure.
In \refscite{Mohanta,Fidkowski1,Fidkowski2,Kim}, Majorana physics
is predicted to arise from the coexistence of magnetism and superconductivity.
This means that (physical, spin-$1/2$) time-reversal symmetry is
broken, whereas the $SC^{+-}$-state respects time-reversal symmetry.

\section{Competing Phases and Spin Textures}
\begin{figure*}[tb]
\begin{center}
\includegraphics[width=0.76\linewidth]{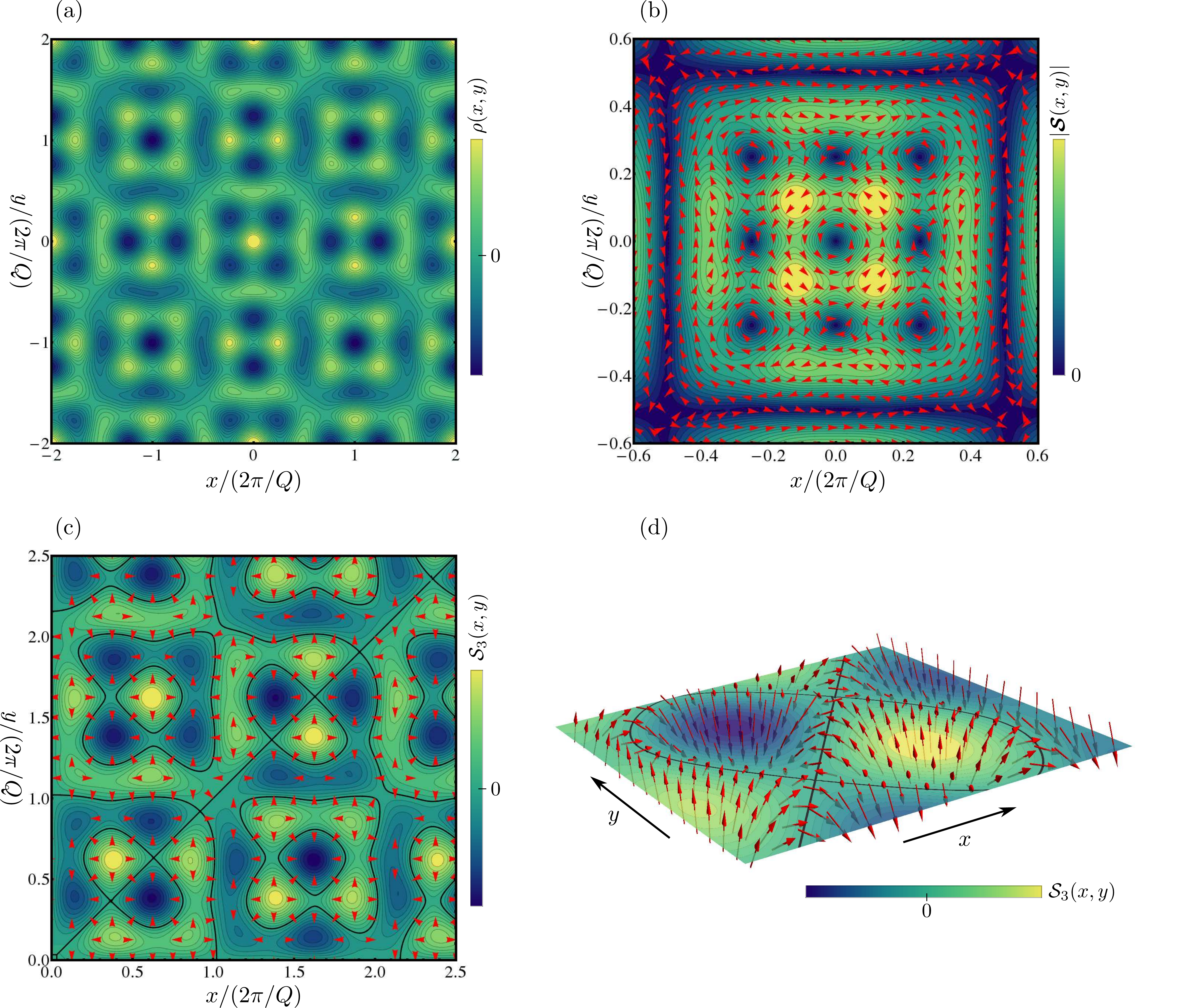}
\caption{Illustration of the spatial structure of the different density wave phases in \figref{RGflowAndPhaseDiagram}(c) and (d) using a nesting vector $(2,0.48)Q$. Part (a) shows the charge density pattern of $CDW^{12}$. In $SDW^{12}$, where the nesting vector is again given by $\vec{Q}_{12}=\vec{k}_1 + \vec{k}_2$, the spin lies approximately in the $xy$-plane. As shown in (b) using red arrows to indicate the direction of the spin, one finds a lattice of vortices. In case of $SDW^{11}$, the nesting vector is $2\vec{k}_1$ and we observe a lattice of Skyrmions and Antiskyrmions as illustrated in (c), where the red arrows indicate the direction of the $xy$-components of the spin and the black lines are the zeros of $\mathcal{S}_3$. Part (d) illustrates one of the emerging closely bound Skyrmion-Antiskyrmion pairs. The texture of $SDW^{22}$ (nesting vector $2\vec{k}_2$) is identical to $SDW^{11}$ upon replacing $\mathcal{S}_3 \rightarrow -\mathcal{S}_3$.}
\label{IllustrationOfDWPhases}
\end{center}
\end{figure*}

Eventually, our RG flow will always favor a superconducting
state. However, by successively reducing the characteristic energy scale, we are increasingly sensitive
to details of the low-energy theory and, consequently, the fact that
the nesting is not perfect for $m_{l}/m_{h}>0$ becomes
relevant. In this sense, any finite $m_{l}/m_{h}$ introduces a cutoff
to the flow. If the flow is cut off before the superconducting instabilities
take place, other competing phases can emerge, as illustrated in \figref{RGflowAndPhaseDiagram}(b).
Depending on the values of the non-flowing coupling constants $u_{00}$
and $u_{30}$, one can either find a charge density wave ($CDW^{12}$),
three different spin density waves ($SDW^{11}$, $SDW^{22}$, $SDW^{12}$)
or the corresponding superconducting states are dominant for arbitrary
$m_{l}/m_{h}$ as shown in \figref{RGflowAndPhaseDiagram}(c) and
(d). The superscripts in the density waves $CDW^{ij}$ and $SDW^{ij}$
refer to the particle-hole expectation value $\Delta_{(-,i),(+,j)}^{\text{DW}}$
(and $i\leftrightarrow j$ if $i\neq j$) that is non-zero in the
respective phase. The difference between $CDW^{12}$ and $SDW^{12}$
is the relative sign of $\Delta_{(-,i),(+,j)}^{\text{DW}}$ and $\Delta_{(-,j),(+,i)}^{\text{DW}}$,
rendering the order parameter symmetric and antisymmetric under time-reversal
in the former and in the latter case, respectively.

The spatial structure of the charge and spin density waves can easily
be determined from the wave functions of the system and the
order parameters $\Delta_{\alpha,\beta}^{\text{DW}}$. As in the case
of the superconducting order parameter, we assume that no additional point group symmetry is broken. In the case of
the $CDW^{12}$-phase, one then finds that the local charge density
is given by 
\begin{equation}
\rho(\vec{x})\propto\cos\left(\vec{Q}_{12}\cdot\vec{x}\right)+\dots,\label{CDWStructure}
\end{equation}
where $\vec{Q}_{12}=\vec{k}_{1}+\vec{k}_{2}$ is the associated nesting
vector. The first contribution stems solely from the nested subspace,
highlighted in red in \figref{SpectrumAndWavefunctions}(a) and the
ellipsis stands for the terms emanating from the remaining three subspaces
which are fully determined by the $\pi/2$-rotation and reflection
symmetry at the $xz$-axis. The resulting charge profile is illustrated
in \figref{IllustrationOfDWPhases}(a).
Note that the periodicity crucially depends on the ratio of the $x$- and $y$-component
of the nesting vector $\vec{Q}_{12}$.

Similarly, the spatial structure of the spin density waves $SDW^{12}$
and $SDW^{11}$, $SDW^{22}$ can be calculated (for details see \supl)
yielding the textures shown in \figref{IllustrationOfDWPhases}(b)
and (c), respectively. 
Here we have used that, in the red regions of \figref{SpectrumAndWavefunctions}(a),
the spins are approximately aligned along the $y$-axis (see \figref{SpectrumAndWavefunctions}(b)
and (c)). Within this approximation, the expectation value of the
spin lies in the $xy$-plane in case of the spin density phase $SDW^{12}$. 
The two-dimensional vector field is therefore a lattice
of vortices both with positive and negative winding number. In the
phases $SDW^{11}$ and $SDW^{22}$, the spin is free to rotate in
three dimensions. One finds a complicated periodic arrangement of
isolated Skyrmions and Antiskyrmions as well as closely bound Skyrmion-Antiskyrmion
pairs (see \figref{IllustrationOfDWPhases}(d)). The emergence of
a Skyrmion lattice, which leads to interesting physical
effects (see e.g.~\refcite{Rosch}), is consistent with recent work\cite{Li2014,Garaud}
pointing out that these magnetic topological defects naturally appear as solutions
of the Ginzburg Landau equations for systems with spin-orbit interaction.

On top of that, the difference between the density wave phases in
\figref{RGflowAndPhaseDiagram}(c) and (d) neighboring the superconducting
states $SC^{+-}$ and $SC^{++}$ can be exploited to gain information
about the pairing mechanism in the heterostructure. As the orbital contribution to the magnetization is negligible for large mass anisotropies, the experimental
observation of in-plane magnetization\cite{Li2011} is only consistent
with the $SDW^{12}$-state. This implies that the superconducting
phase of SrTiO$_{3}$/LaAlO$_{3}$ is supposed to be unconventional
and topologically nontrivial. 

As already stated above, we have also considered the case of different
Fermi velocities, $v_{1}\neq v_{2}$ (see \supl~for more details
of the analysis). Then all four backscattering couplings flow. Nonetheless,
exactly as before, the leading instability is generically superconducting
for sufficiently large mass anisotropies. However, in the present
case, the anomalous expectation value $\Delta_{(-,j),(+,j)}^{\text{SC}}$ is only
finite on the Fermi surface with the larger Fermi velocity. Remarkably,
we still find that the superconductor resulting from the conventional
electron-phonon pairing mechanism is topologically trivial, whereas
the unconventional superconductor is nontrivial. This proves that
the correspondence between the pairing mechanism and the topological
properties of the superconducting phases in the heterostructure holds
irrespective of the values of the Fermi velocities. For completeness,
we also considered the case of very weak spin-orbit interaction where
the energetic cutoff of the low-energy model is much larger than the
spin-orbit splitting. Then the red regions in \figref{SpectrumAndWavefunctions}(a)
overlap pairwise and, consequently, momentum conservation is much
less restrictive making more backscattering terms possible. Surprisingly,
still in this situation, the observation of a topologically nontrivial
superconducting phase is only consistent with the pairing mechanism
being unconventional.

The phase diagram of the two-dimensional electron fluid that forms
at the interface between the perovskite oxides LaAlO$_{3}$ and SrTiO$_{3}$
combines two fascinating notions of condensed matter physics: Topology
and unconventional superconductivity. We find that, very generically,
the observation of signatures of topologically nontrivial superconductivity,
such as the appearance of Majorana bond states, directly implies that
the underlying pairing mechanism must be unconventional. In addition,
the spin density wave phases competing with topological superconductivity
show topological spatial textures as well. Depending on the value
of the coupling constants, we find lattices of both Skyrmions and
vortices. 


\vspace{1.5em}
\textbf{Acknowledgements}. -- We are grateful for discussions with S.~Beyl, A.~V.~Chubukov, A.~M.~Finkel'stein, E.~J.~König, D.~Mendler, and A.~D.~Mirlin. We acknowledge financial support by the Deutsche Forschungsgemeinschaft through grant SCHM 1031/4-1.
 
\newpage
\section{Supplementary information}
\subsection{General symmetry analysis}
The symmetry classification of the electron-electron interaction can be performed efficiently by introducing a specific phase convention for the local eigenbasis of the free Hamiltonian. Here we define this convention which will then be used to represent the point symmetries and time-reversal on the helicity operators $c$, $c^\dagger$. Finally, all possible momentum independent interaction terms within the most strongly nested subspaces (see \figref{FermiSurfaceConventions}(a)) will be derived. We consider all three relevant cases, non-overlapping low-energy subspaces with both identical and different Fermi velocities as well as quasi-degenerate Fermi surfaces (see \figref{FermiSurfaceConventions}(b)-(d)), simultaneously.

\subsubsection{Phase convention and representation of symmetries}
Using a path-integral representation, the quadratic part of the theory can be written as
\begin{equation}
 S_0 = T\sum_{\omega_n}\sum_{\vec{k}} \overline{\Psi}_\alpha(k) \left[-i\omega\delta_{\alpha,\beta} + H_{\alpha,\beta}(\vec{k})\right] \Psi_\beta(k), \label{nondiagonalizedquadraticaction}
\end{equation}  
where $k\equiv(\omega_n,\vec{k})$ and $\Psi$, $\overline{\Psi}$ are four-component Grassmann fields describing spinful Fermions in the two orbitals $\{xz,yz\}$. Furthermore, $H$ is the Hamiltonian defined in the main text characterized by the anisotropic masses (\ref{eq:disp}) and the spin-orbit coupling in \equref{eq:spinorbfin}.

\begin{figure}[bt]
\begin{center}
\includegraphics[width=0.7\linewidth]{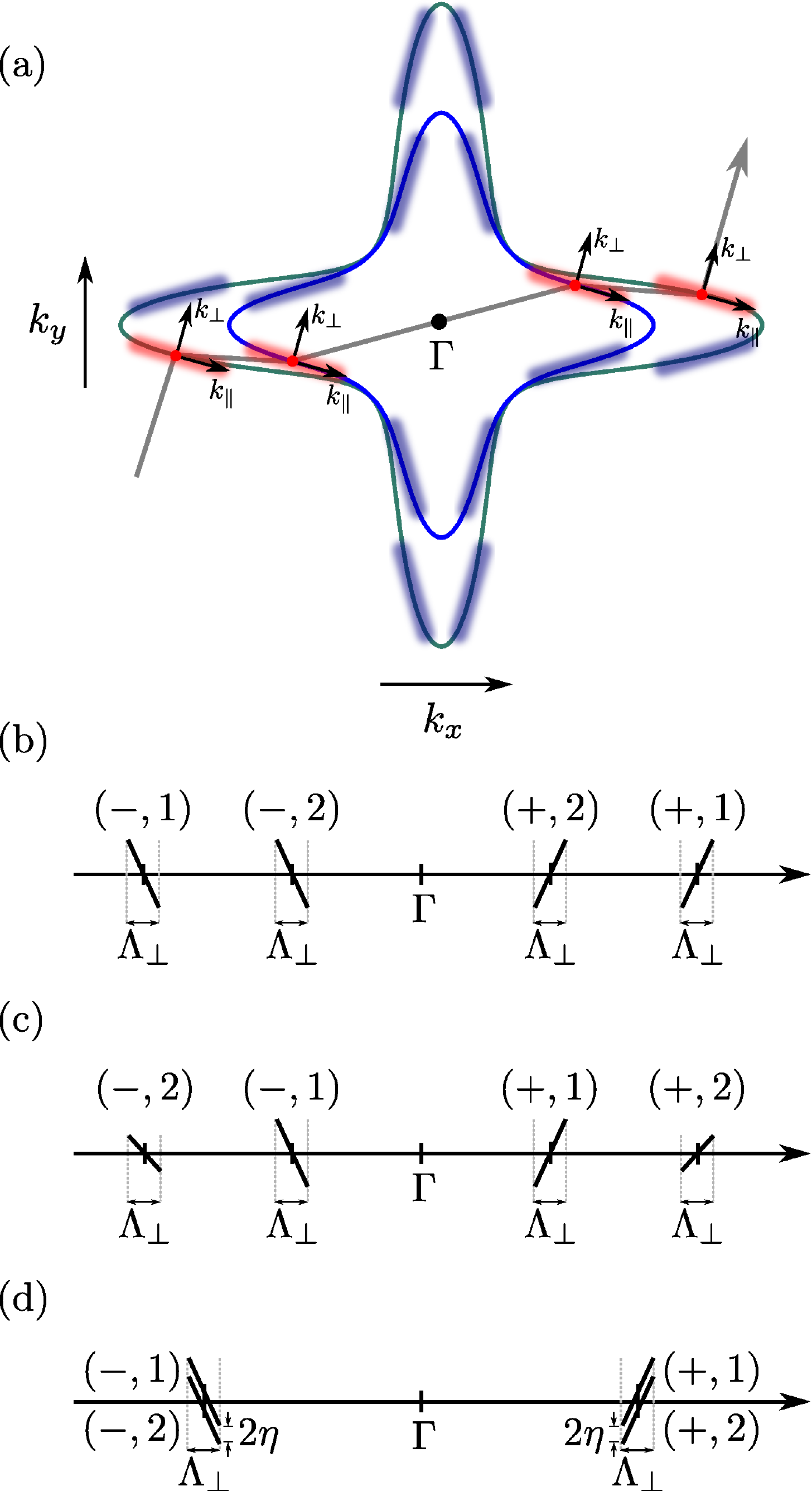}
\caption{Illustration of the low-energy description. In (a), the Fermi surface is shown and the strongly nested subspaces as well as the local coordinate systems of the low-energy theory are indicated. The spectrum along the gray arrow is shown schematically in part (b) and (c) for strong spin-orbit coupling in case of identical and different Fermi velocities, respectively, and in (d) for quasi-degenerate Fermi surfaces ($\eta\ll v_j\Lambda_\perp$).}
\label{FermiSurfaceConventions}
\end{center}
\end{figure}

We diagonalize $S_0$ by performing the unitary transformation
\begin{equation}
 \Psi_{\alpha}(k) = \mathcal{U}_{\alpha,\alpha'}(\vec{k}) f_{\alpha'}(k), \,\, \overline{\Psi}_{\alpha}(k) = \mathcal{U}^*_{\alpha,\alpha'}(\vec{k}) \bar{f}_{\alpha'}(k), \label{IntroductionOfcs}
\end{equation} 
where
\begin{equation}
 \mathcal{U}(\vec{k}) =[\phi_1(\vec{k}),\phi_2(\vec{k}),\phi_3(\vec{k}),\phi_4(\vec{k})] \label{Uconcrete}
\end{equation} 
with $\phi_\alpha(\vec{k})$ denoting an eigenvector of $H(\vec{k})$. As explained in the main text, we can restrict the analysis of instabilities to one of the most strongly nested subspaces. We choose the subspace highlighted in red in \figref{FermiSurfaceConventions}(a) and introduce helicity fields $c_{(\sigma,j)}$ and $\bar{c}_{(\sigma,j)}$ in the local coordinate systems yielding 
\begin{equation}
 S_0 =  \int_k \bar{c}_{(\sigma,j)}(k) \left[-i\omega_n + \sigma v_{j} k_\perp + s_j \eta \right] c_{(\sigma,j)}(k) \label{NoninteractingAction}
\end{equation}
after linearizing the spectrum. Here $s_1=+1$, $s_2=-1$ and $\eta$ denotes the spin-orbit splitting in the case of quasi-degenerate Fermi surfaces. For stronger spin-orbit coupling, where the four red regions in \figref{FermiSurfaceConventions}(a) are disjoint, one has $\eta=0$ by construction. In \equref{NoninteractingAction} and in the following, we use the compact notation $k\equiv(\omega_n,k_\parallel,k_\perp)$ and

\begin{equation}
 \int_k \dots = \int_{-\infty}^{\infty}\frac{\diff \omega}{2\pi} \int_{-\Lambda_\parallel}^{\Lambda_\parallel} \frac{\diff k_\parallel}{2\pi} \int_{-\Lambda_\perp}^{\Lambda_\perp} \frac{\diff k_\perp}{2\pi} \dots,
\end{equation} 
where $\Lambda_\perp$ and $\Lambda_\parallel$ are the momentum cutoffs normal and tangential to the Fermi surface.
If the Fermi velocities are identical, we will use the notation introduced in the main text where $j=1$ ($j=2$) refers to the outer (inner) Fermi surface. If this is not the case, it will be most convenient to label the fields such that $v_1 > v_2$.

To make the helicity operators unique, we have to fix the phases of the eigenstates in \equref{Uconcrete}. This is achieved by exploiting the invariance of the Hamiltonian under $\pi$-rotation $R^{c2}$ and time-reversal $\Theta$. The former symmetry implies that 
\begin{equation}
 H(\vec{k}) = \mathcal{R}^{c2}_\Psi H(-\vec{k}) {\mathcal{R}^{c2}_\Psi}^\dagger, \quad \mathcal{R}^{c2}_\Psi = i\sigma_z,
\end{equation} 
and hence we can construct the eigenstates with negative $k_x$ from those with $k_x>0$ via
\begin{equation}
 \phi_\alpha(-\vec{k}) := {\mathcal{R}_\Psi^{c2}}^\dagger\phi_\alpha(\vec{k}), \qquad k_x>0. \label{PhaseConvention1}
\end{equation}  

Consecutive application of time-reversal and $\pi$-rotation leads to the $\vec{k}$-space local antiunitary symmetry 
\begin{equation}
 H(\vec{k}) = \sigma_x H^*(\vec{k})\sigma_x
\end{equation} 
of the Hamiltonian. If the Fermi surfaces in \figref{FermiSurfaceConventions} are non-degenerate, we can adjust the phases of the eigenstates such that 
\begin{equation}
 \phi_\alpha(\vec{k}) =  \sigma_x\phi_\alpha^*(\vec{k}) \label{PhaseConvention2}
\end{equation} 
for $k_x>0$. From \equref{PhaseConvention1}, it follows that \equref{PhaseConvention2} actually holds also for $k_x<0$. 
In addition, we have shown that \equref{PhaseConvention2} can still be satisfied if the Fermi surfaces are exactly degenerate. 

Having fixed the phases of the local eigenstates, the representation of time-reversal and $\pi$-rotation symmetry on the helicity fields is well defined. Note that the remaining elements of the point group $C_{4v}$ cannot be represented in the most strongly nested subspace as these operations act between different subspaces. For the very same reason, however, the remaining symmetries are also irrelevant when deriving the most general interaction within one the subspaces.

Time-reversal acts according to
\begin{subequations}
\begin{align}
 \Psi_\alpha(\omega,\vec{k}) \, &\stackrel{\Theta}{\longrightarrow} \, \left(i\sigma_y\right)_{\alpha,\beta} \overline{\Psi}_\beta(\omega,-\vec{k}), \\
 \overline{\Psi}_\alpha(\omega,\vec{k}) \, &\stackrel{\Theta}{\longrightarrow} \, \Psi_\beta(\omega,-\vec{k}) \left(i\sigma_y\right)_{\beta,\alpha}
\end{align}\end{subequations}
in the basis of \equref{nondiagonalizedquadraticaction} and, consequently, as
\begin{subequations}
\begin{align}
  f_\alpha(k) &\,\stackrel{\Theta}{\longrightarrow}\, i\left[\mathcal{U}^\dagger(\vec{k})  \sigma_y \, \mathcal{U}^*(-\vec{k}) \right]_{\alpha,\alpha'} \bar{f}_{\alpha'}(\omega,-\vec{k}) \\
 \bar{f}_\alpha(k) &\,\stackrel{\Theta}{\longrightarrow}\,  if_{\alpha'}(\omega,-\vec{k}) \left[\left( \mathcal{U}^\dagger(\vec{k}) \sigma_y   \mathcal{U}^*(-\vec{k})\right)^\dagger\right]_{\alpha',\alpha}
\end{align}\end{subequations}
in the local eigenbasis. Using \equsref{PhaseConvention1}{PhaseConvention2}, we can write
\begin{align}
 \mathcal{U}^\dagger(\vec{k})  \sigma_y \, \mathcal{U}^*(-\vec{k}) &= \mathcal{U}^\dagger(\vec{k})  \sigma_y \left[\phi_1^*(-\vec{k}),\dots\right] \nonumber \\
&=\mathcal{U}^\dagger(\vec{k})  \sigma_y\sigma_x \left[\phi_1(-\vec{k}),\dots\right]  \nonumber \\
&=-\sign(k_x) \mathcal{U}^\dagger(\vec{k})  \sigma_z \sigma_z \left[\phi_1(\vec{k}),\dots\right]  \nonumber \\
&= -\sign(k_x)\mathbbm{1}
\end{align}
and, thus, conclude
\begin{subequations}
\begin{align}
 c_{(\pm,j)}(\omega,k_\parallel,k_\perp) \, &\stackrel{\Theta}{\longrightarrow} \, \mp i \bar{c}_{(\mp,j)}(\omega,-k_\parallel,-k_\perp), \\
 \bar{c}_{(\pm,j)}(\omega,k_\parallel,k_\perp) \, &\stackrel{\Theta}{\longrightarrow} \, \mp i c_{(\mp,j)}(\omega,-k_\parallel,-k_\perp).
\end{align}\label{TRTransformation}\end{subequations}

Similarly, for the $\pi$-rotation symmetry, one finds
\begin{align}
 c_{(\pm,j_\alpha)}(\omega,k_\parallel,k_\perp) \, &\stackrel{R^{c2}}{\longrightarrow} \, \mp c_{(\mp,j_\alpha)}(\omega,-k_\parallel,-k_\perp) \label{PiRotationRepresentation}
\end{align}
and the same for $\bar{c}$.

\subsubsection{Symmetry analysis of the interaction}
Now we will derive the most general momentum independent interaction of the low-energy theory consistent with the symmetries of the system. Let us write 
\begin{align}
\begin{split}
 S_{\text{int}} =\int_{k_1,k_2,k_3,k_4}  &\bar{c}_\alpha(k_4)\bar{c}_\beta(k_3)c_\gamma(k_2)c_\delta(k_1) \,\mathcal{W}^{\alpha\beta}_{\gamma\delta}  \\ 
&\qquad\times\delta(k_1+k_2-k_3-k_4), \label{GeneralProjectedInteraction}\end{split}
\end{align}
where the Greek letters are double indices comprising $\sigma=\pm$ and $j=1,2$.
The tensor $\mathcal{W}$ has to satisfy
\begin{equation}
 \mathcal{W}^{\alpha\beta}_{\gamma\delta} = \left(\mathcal{W}^{\delta\gamma}_{\beta\alpha}\right)^*\label{WHermitictiy}
\end{equation} 
due to Hermiticity and, as a consequence of Fermi statistics, can be chosen such that
\begin{equation}
 \mathcal{W}^{\alpha\beta}_{\gamma\delta} = -\mathcal{W}^{\beta\alpha}_{\gamma\delta} = -\mathcal{W}^{\alpha\beta}_{\delta\gamma}. \label{AntiSymmetryOfLowEnergyInteractionTensor}
\end{equation} 
It turns out that the dimensionless parameterization,
\begin{equation}
 \omega^{\alpha\beta}_{\gamma\delta} = \frac{\Lambda_{\parallel}}{2\pi^2v_1}\mathcal{W}^{\alpha\beta}_{\gamma\delta} \label{DimensionlessNotation}
\end{equation} 
with
\begin{align}
\begin{split}
&\omega^{(\sigma_\alpha,j_\alpha)(\sigma_\beta,j_\beta)}_{(\sigma_\gamma,j_\gamma)(\sigma_\delta,j_\delta)} = \\  
&= \begin{cases} V^{j_\alpha,j_\beta}_{j_\gamma,j_\delta}(\sigma) , & \sigma_\alpha=\sigma_\beta=\sigma_\gamma=\sigma_\delta=\sigma, \\ W^{j_\alpha,j_\beta}_{j_\gamma,j_\delta}, & (\sigma_\alpha,\sigma_\beta,\sigma_\gamma,\sigma_\delta) = (-,+,+,-), \\ W^{j_\beta,j_\alpha}_{j_\delta,j_\gamma}, & (\sigma_\alpha,\sigma_\beta,\sigma_\gamma,\sigma_\delta) = (+,-,-,+), \\ -W^{j_\beta,j_\alpha}_{j_\gamma,j_\delta}, & (\sigma_\alpha,\sigma_\beta,\sigma_\gamma,\sigma_\delta) = (+,-,+,-), \\ -W^{j_\alpha,j_\beta}_{j_\delta,j_\gamma}, & (\sigma_\alpha,\sigma_\beta,\sigma_\gamma,\sigma_\delta) = (-,+,-,
+), \\ 0, & \text{otherwise}, \label{InteractionTensorGeneral} \end{cases} \end{split}
\end{align}
is very convenient for the following analysis. In \equref{InteractionTensorGeneral}, we have already taken into account \equref{AntiSymmetryOfLowEnergyInteractionTensor} and that only forward scattering (described by $V$) and backscattering ($W$) are allowed by momentum conservation, which is directly clear from \figref{FermiSurfaceConventions}. Throughout this work, we assume that Umklapp processes are not possible.
Due to Fermi statistics, the forward scattering tensors must have the form
\begin{equation}
 V^{j_\alpha, j_\beta}_{j_\gamma, j_\delta}(\sigma) = g_0(\sigma) \left[\delta_{j_\alpha,j_\delta}\delta_{j_\beta,j_\gamma}-\delta_{j_\alpha,j_\gamma}\delta_{j_\beta,j_\delta}\right],
\end{equation} 
whereas the backscattering tensor has $16$ degrees of freedom, which we parametrize according to
\begin{equation}
  W^{j_\alpha,j_\beta}_{j_\gamma,j_\delta} = \sum_{s,s'=0}^3 g_{ss'} \left(\sigma_s\right)_{j_\alpha,j_\delta} \left(\sigma_{s'}\right)_{j_\beta,j_\gamma}. \label{ConstantWexpansion}
\end{equation}
The Hermiticity constraint in \equref{WHermitictiy} implies that $g_0(\sigma), \, g_{ss'} \in \mathbbm{R}$. Note that $g_{ss'} \propto u_{ss'}$ with $u_{ss'}$ used in the main text to define the backscattering terms.

Next, let us derive the constraints resulting from $\pi$-rotation symmetry. Demanding that \equref{GeneralProjectedInteraction} be invariant under \equref{PiRotationRepresentation}, we find 
\begin{subequations}
\begin{align}
 V^{j_\alpha,j_\beta}_{j_\gamma,j_\delta}(+) &\stackrel{!}{=}V^{j_\alpha,j_\beta}_{j_\gamma,j_\delta}(-), \\
 W^{j_\beta,j_\alpha}_{j_\delta,j_\gamma} &\stackrel{!}{=} W^{j_\alpha,j_\beta}_{j_\gamma,j_\delta}.
\end{align}\end{subequations}
The former conditions means that, as expected, forward scattering is identical for the patches centered around $\vec{k}_j$ and $-\vec{k}_j$. Consequently, all forward scattering processes are characterized by one coupling constant $g_0\equiv g_0(+)=g_0(-)$. Applying the expansion (\ref{ConstantWexpansion}), the second constraint is equivalent to $g^T=g$ as stated in the main text.

Similarly, to make the interaction time-reversal symmetric, we require invariance of \equref{GeneralProjectedInteraction} under \equref{TRTransformation}. Again using the parameterization (\ref{InteractionTensorGeneral}), we find that $V$ is not further restricted, whereas the backscattering tensor has to satisfy
\begin{equation}
 W^{j_\delta,j_\gamma}_{j_\beta,j_\alpha} \stackrel{!}{=} W^{j_\alpha,j_\beta}_{j_\gamma,j_\delta}.
\end{equation} 
In the representation (\ref{ConstantWexpansion}) this is equivalent to demanding $g_{s,s'}=0$ if either $s=2$ or $s'=2$.

\begin{figure}[b]
\begin{center}
\includegraphics[width=\linewidth]{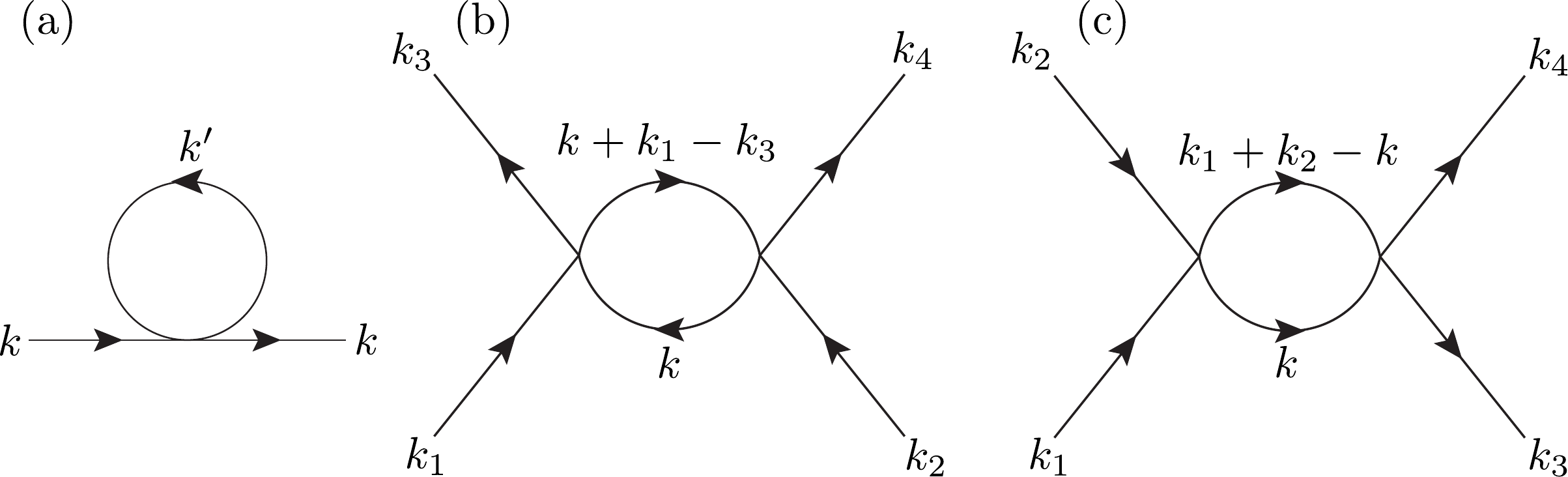}
\caption{Diagrams to be evaluated for the RG. Closed loops involve integration over fast modes only.}
\label{OneLoopDiagrams}
\end{center}
\end{figure}

Consequently, in the limit of weak spin-orbit interaction, where $0<\eta\ll v_j\Lambda_\perp$ and the red regions in \figref{FermiSurfaceConventions} overlap pairwise, the backscattering tensor is given by \equref{ConstantWexpansion} with   
\begin{equation}
 g=\begin{pmatrix} g_{00} & g_{10} & 0 & g_{30} \\ g_{10} & g_{11} & 0 & g_{31} \\ 0 & 0 & g_{22} & 0 \\ g_{30} & g_{31} & 0 & g_{33}  \end{pmatrix}. \label{GeneralgOverlapping}
\end{equation} 

However, if the four most strongly nested subspaces are disjoint, momentum conservation rules out further backscattering terms. Writing down all interaction terms that are consistent with momentum conservation and expanding them in Pauli matrices as in \equref{ConstantWexpansion}, one finds that only $g_{00}$, $g_{11}=-g_{22}$, $g_{33}$, $g_{21}=g_{12}$, $g_{30}$ and $g_{03}$ can be finite. Comparison with \equref{GeneralgOverlapping} then yields
\begin{equation}
 g=\begin{pmatrix} g_{00} & 0 & 0 & g_{30} \\ 0 & g_{11} & 0 & 0 \\ 0 & 0 & -g_{11} & 0 \\ g_{30} & 0 & 0 & g_{33}  \end{pmatrix}. \label{GeneralgDisjoint}
\end{equation}

\subsection{Wilson RG}
In this part, we provide more details of the RG calculation and discuss the flow equations for all three regimes in \figref{FermiSurfaceConventions}(b)-(d).
\subsubsection{Generic form of the RG equations}
In the Wilson approach, applied to Fermions with a finite Fermi surface in \refcite{Shankar}, fast modes with momenta $\Lambda_\perp e^{-\Delta l}<k_\perp<\Lambda_\perp$, $\Delta l>0$, are integrated out yielding, after proper rescaling, an effective action with renormalized parameters. The quadratic part of the action simply splits into the contributions from the fast and slow modes, whereas the interaction leads to nontrivial terms in the effective action that can only be treated perturbatively. 

The corresponding one-loop contributions are shown diagrammatically in \figref{OneLoopDiagrams}. The tadpole diagram, \figref{OneLoopDiagrams}(a), represents the impact of the interaction on the bands of the system. Here and in the following, we will neglect this contribution to the RG flow, since, by definition, we assume that all possible interaction effects on the chemical potential and on the spin-orbit coupling have already been accounted for by $S_0$.

The other two diagrams, \figref{OneLoopDiagrams}(b) and (c), are usually referred to as ZS and BCS, respectively, and lead to the corrections
\begin{equation}
 \Delta_{\text{BCS}}\mathcal{W}_{\gamma\delta}^{\alpha\beta} = 2 \mathcal{W}^{\alpha\beta}_{\mu\nu}\mathcal{W}^{\mu\nu}_{\gamma\delta} \int_k G^>_{\mu}(k)G^>_{\nu}(k_1+k_2-k) \label{BCSshell}
\end{equation}
and
\begin{align}
\begin{split}
 \Delta_{\text{ZS}}\mathcal{W}_{\gamma\delta}^{\alpha\beta} &= -4 \left(\mathcal{W}^{\alpha\nu}_{\gamma\mu}\mathcal{W}^{\mu\beta}_{\nu\delta} - \mathcal{W}^{\beta\nu}_{\gamma\mu}\mathcal{W}^{\mu\alpha}_{\nu\delta} \right) \\
&\qquad\times\int_k G^>_{\mu}(k+k_1-k_3) G^>_{\nu}(k) \label{ZSshell}
\end{split}
\end{align}
of the interaction tensor $\mathcal{W}$. In \equsref{BCSshell}{ZSshell}, we have introduced the Green's function
\begin{equation}
 G^>_\alpha(k) = \frac{\theta(|k_\perp|-\Lambda_{\perp}e^{-\Delta l})\theta(\Lambda_{\perp}-|k_\perp|)}{i\omega-\sigma_\alpha v_{j_\alpha} k_\perp-s_j\eta} \label{ShellGreensFunction} 
\end{equation} 
of fast modes. Note that $\Delta_{\text{BCS}}\mathcal{W}_{\gamma\delta}^{\alpha\beta}$ and $\Delta_{\text{ZS}}\mathcal{W}_{\gamma\delta}^{\alpha\beta}$ have been symmetrized to satisfy \equref{AntiSymmetryOfLowEnergyInteractionTensor}. Evaluating the shell integrals asymptotically in the limit $\Delta l \rightarrow 0$ and using the dimensionless parameterization (\ref{DimensionlessNotation}), one finds the tensor valued RG equation
\begin{widetext}
\begin{align}
\begin{split}
 \der{\omega_{\gamma\delta}^{\alpha\beta}(l)}{l} &=  \left(1-\delta_{\sigma_\mu,\sigma_\nu}\right) \Biggl[ \omega^{\alpha\beta}_{\mu\nu}(l)\omega^{\mu\nu}_{\gamma\delta}(l) \sum_{p=+,-} \frac{1}{x_{j_\mu}(1+p\,\kappa_{j_\mu}) + x_{j_\nu}(1 +p\, \kappa_{j_\nu})} \\
&\qquad + 2\left(\omega^{\alpha\nu}_{\gamma\mu}(l)\omega^{\mu\beta}_{\nu\delta}(l) - \omega^{\beta\nu}_{\gamma\mu}(l)\omega^{\mu\alpha}_{\nu\delta}(l)\right)\sum_{p=+,-} \frac{1}{x_{j_\mu}(1+p\,\kappa_{j_\mu}) + x_{j_\nu}(1 -p\, \kappa_{j_\nu})} \Biggr],
\label{TensorValuedRGequations} \end{split}
\end{align}
\end{widetext}
where $x_j:=v_j/v_1$ and $\kappa_{j}:=s_j\eta/(\Lambda_{\perp} v_j)$ have been defined. 
From \equsref{InteractionTensorGeneral}{TensorValuedRGequations}, it is already clear that
\begin{equation}
 \der{\omega^{(\sigma,j_\alpha)(\sigma,j_\beta)}_{(\sigma,j_\gamma)(\sigma,j_\delta)}}{l} = \der{V^{j_\alpha,j_\beta}_{j_\gamma,j_\delta}(\sigma)}{l} = 0, \label{NoFlowForward}
\end{equation} 
i.e., irrespective of the Fermi velocities and the strength of the spin-orbit coupling, the forward scattering terms are not renormalized.

To simplify the following analysis, let us set $\eta \rightarrow 0$ in the flow equation (\ref{TensorValuedRGequations}). Note that this rules out only the intermediate regime where the energetic cutoff is of the same order as the spin-orbit splitting $\eta$, since, for stronger spin-orbit interaction, we have $\eta=0$ by construction (see \figref{FermiSurfaceConventions}(b) and (c)).
Inserting $(\sigma_\alpha,\sigma_\beta,\sigma_\gamma,\sigma_\delta)=(-,+,+,-)$ in \equref{TensorValuedRGequations} then yields, after some algebra, the flow equation
\begin{align}
\begin{split}
 \der{W^{j_\alpha,j_\beta}_{j_\gamma,j_\delta}}{l} = \sum_{j_\mu,j_\nu}& \frac{4}{x_{j_\mu}+x_{j_\nu}}\biggl(W^{j_\alpha,j_\nu}_{j_\gamma,j_\mu}W^{j_\mu,j_\beta}_{j_\nu,j_\delta} \\
&-W^{j_\alpha,j_\beta}_{j_\nu,j_\mu}W^{j_\mu,j_\nu}_{j_\gamma,j_\delta}\biggr)\end{split} \label{WTensorFlowEquation}
\end{align} 
of the backscattering tensor. Here the contribution of the first and second line emanate from the ZS and BCS diagram, respectively.

Next, we will restate \equref{WTensorFlowEquation} in terms of the coupling constants $g_{ss'}$ for the two cases of large spin-orbit coupling and quasi-degenerate Fermi surfaces. 

\begin{figure*}[t]
\begin{center}
\includegraphics[width=\linewidth]{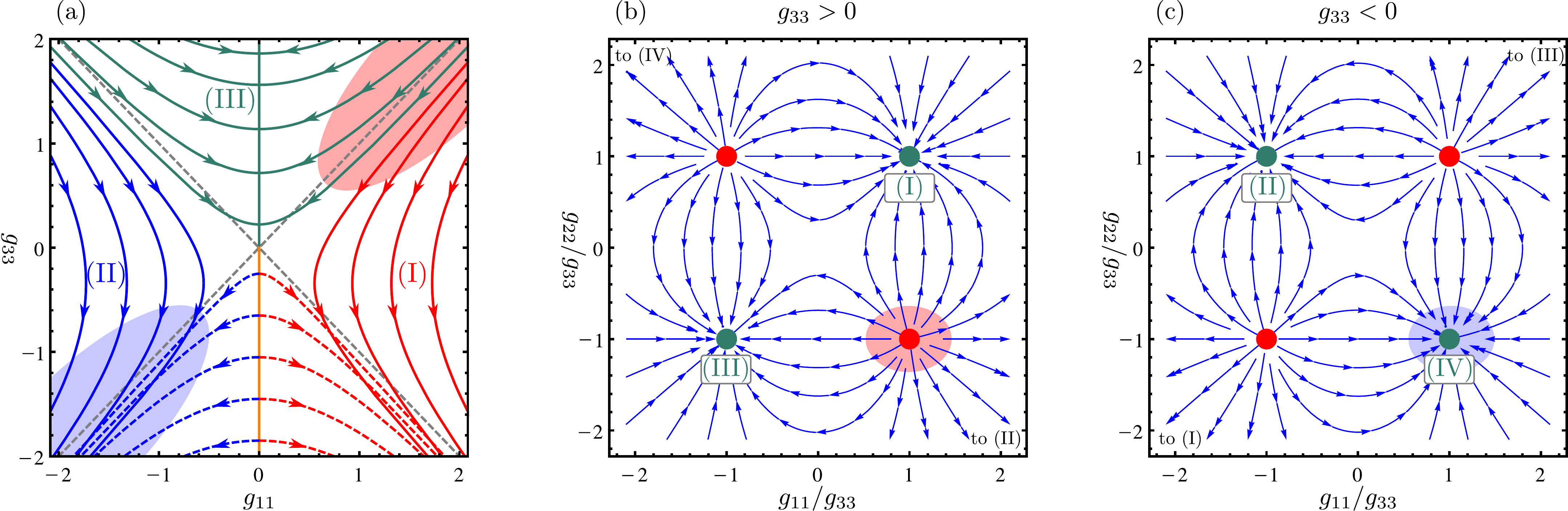}
\caption{RG flow in the cases that have not been discussed in the main text. Part (a) shows the projection of the flow for $v_2/v_1 = 0.4$ using $g_{00}=1$ and $g_{30}=0.1$. For $g_{00}<0$, the projection has essentially the same structure. In (b) and (c), the reduced flow in cases of quasi-degenerate Fermi surfaces is shown for $g_{33}>0$ and $g_{33}<0$, respectively. In all three plots, the red (blue) regions correspond to an initially repulsive (attractive) interaction thus identifying the unconventional (conventional) superconductor.}
\label{AdditionalRGFlows}
\end{center}
\end{figure*}

\subsubsection{Large spin-orbit coupling}
To begin with the former, we insert the parameterization (\ref{ConstantWexpansion}) using $g_{ss'}$ as given in \equref{GeneralgDisjoint} into \equref{WTensorFlowEquation} and find

\begin{subequations}
 \begin{align}
  \der{g_{00}}{l} &= -2g_{11}^2\frac{(v_1-v_2)^2}{v_2\left( v_1 + v_2 \right)}, \\
  \der{g_{30}}{l} &= -2g_{11}^2\frac{v_1-v_2}{v_2}, \\
  \begin{split}
\der{g_{11}}{l} &= -\frac{2g_{11}}{v_2\left( v_1 + v_2 \right)}\Bigl[g_{33}(v_1^2+v_2^2+6v_1 v_2) \\ 
& \qquad + g_{00}(v_1-v_2)^2 - 2g_{30}\left( v_1^2-v_2^2 \right) \Bigr], \end{split} \\
  \der{g_{33}}{l} &= -2g^2_{11}\frac{v_1^2+v_2^2+6v_1 v_2}{v_2\left( v_1 + v_2 \right)}.
 \end{align}\label{RGflowequations2}\end{subequations}
Setting $v_1 \rightarrow v_2$ in \equref{RGflowequations2}, one obtains 
\begin{equation}
  \der{g_{11}}{l} = -8g_{11}g_{33}, \qquad \der{g_{33}}{l} = -8g_{11}^2, \label{RGflowequations1}
\end{equation}
whereas $g_{00}$ and $g_{30}$ do not flow. This is the limit that has been discussed in detail in the main text. The resulting flow is shown in \figref{RGflowAndPhaseDiagram}(a).

If $v_1 \neq v_2$, all four backscattering coupling constants flow. The projection of the RG flow onto the $g_{11}$-$g_{33}$-plane is illustrated in \figref{AdditionalRGFlows}(a). We observe that the structure of the flow diagram is very similar to \figref{RGflowAndPhaseDiagram}(a) and that the three regions (I), (II) and (III) can still be identified. Note that $g_{00}$, $g_{30}$ and $g_{33}$ can only diverge if $g_{11}$ diverges as well which is easily seen from \equref{RGflowequations2}. Hence, none of the couplings diverges in region (III).

\subsubsection{Quasi-degenerate Fermi surfaces}
Finally, we also discuss the situation of very weak spin-orbit coupling where more backscattering terms are possible. To simplify the following analysis, we introduce new Fermion operators $c'$ and $\bar{c}'$ via
\begin{subequations}
\begin{align}
  c(k) &= \begin{pmatrix} e^{i\frac{\vec{\alpha}}{2}\cdot\vec{\sigma}} & 0 \\ 0 & e^{i\frac{\vec{\alpha}}{2}\cdot\vec{\sigma}} \end{pmatrix}c'(k), \qquad \vec{\alpha}\in \mathbbm{R}^3 \\
 \bar{c}(k) &= \bar{c}'(k) \begin{pmatrix} e^{-i\frac{\vec{\alpha}}{2}\cdot\vec{\sigma}} & 0 \\ 0 & e^{-i\frac{\vec{\alpha}}{2}\cdot\vec{\sigma}} \end{pmatrix}, 
\end{align}\end{subequations}
which renders the theory invariant except for a change of the coupling matrix $g_{ss'}$. One can show that, upon properly choosing $\vec{\alpha}$, the coupling matrix $g$ in \equref{GeneralgOverlapping} can be brought into the reduced form
\begin{equation}
 g'=\begin{pmatrix} g_{00} & g'_{10} & 0 & g'_{30} \\ g'_{10} & g'_{11} & 0 & 0 \\ 0 & 0 & g'_{22} & 0 \\ g'_{30} & 0 & 0 & g'_{33}  \end{pmatrix}. \label{GeneralgOverlappingRed}
\end{equation} 
Using this interaction matrix in \equref{WTensorFlowEquation}, we find (neglecting the primes for notational simplicity)
\begin{subequations}
 \begin{align}
  \der{g_{11}}{l} &= 8 g_{22}g_{33}, \\ 
  \der{g_{22}}{l} &= 8 g_{11}g_{33}, \\
  \der{g_{33}}{l} &= 8 g_{11}g_{22}, \\
  \der{g_{p,p'}}{l} &= 0,\qquad\qquad \text{otherwise}.
 \end{align}\label{RGflowequations3}\end{subequations}
The resulting flow of the ratio of the coupling constants is illustrated in \figref{AdditionalRGFlows}(b) and (c) for different signs of $g_{33}$. The reduced flow has four fixed points $((g_{11}/g_{33})^*,(g_{22}/g_{33})^*)=(p_1,p_2)$ with $p_j=+1,-1$ which are stable if and only if $g_{33}p_1p_2>0$. Right at the fixed points, \equref{RGflowequations3} is solved by 
\begin{subequations}
\begin{equation}
 g_{11}(l) = p_1g(l), \quad g_{22}(l) = p_2g(l), \quad g_{33}(l) = g(l)
\end{equation} 
with
\begin{equation}
 g(l)=\frac{g(0)}{1 - 8l\,p_1p_2 g(0)}.
\end{equation} \label{SolutionAtStableFPs}\end{subequations}
Consequently, the coupling constants diverge at all four stable fixed points denoted by (I)-(IV) in \figref{AdditionalRGFlows}(b) and (c).

\subsubsection{Microscopic interaction}
A important part of our analysis is the identification of the pairing mechanisms in the different superconductors. For this purpose, we include matrix elements of the electron-electron interaction between the relevant $d_{xz}$ and $d_{yz}$ orbitals yielding both an intra- ($U$) and inter-orbital ($U'$) Hubbard interaction, a Hund's coupling ($J_H$) term as well as pair-hopping ($J'$). In addition, we use $J=J'$ and $U=U'+2J'$ valid for the usual Coulomb interaction, but our results do not crucially depend on this assumption. 

Projecting the interaction onto the low-energy theory, we find, using the model defined in the main text,
\begin{equation}
 g_{00} \approx g_{11} \approx g_{33}, \quad  |g_{00}|\gg |g_{30}|
\end{equation}
in case of disjoint support in momentum space and
\begin{equation}
 g_{00} \approx g_{11} \approx -g_{22} \approx g_{33}, \quad |g_{00}| \gg |g_{10}|,\, |g_{30}|, \, |g_{31}|
\end{equation} 
for near-spin degeneracy. In this way, we can estimate the initial conditions for the RG flow both for a microscopically repulsive ($g_{00}>0$) and for an electron-phonon induced, attractive ($g_{00}<0$) interaction. The two scenarios correspond, respectively, to the red and blue shaded regions of the flow diagrams in \figref{RGflowAndPhaseDiagram}(a) and \figref{AdditionalRGFlows}. 

\subsection{Mean-field equations and instabilities}
Now we want to investigate which instabilities are associated with the divergences in the RG flow. Following \refsacite{Chubukov}{Vafek}, we analyze the mean-field equations with the renormalized couplings for any instability possible at finite temperature. The leading instability is the one with the highest transition temperature. 

Let us assume spatial and temporal homogeneity of the particle-hole,
\begin{equation}
 \Delta_{\alpha,\beta}^{\text{DW}} = \frac{1}{\beta V} \int_k\braket{\bar{c}_\alpha(k)c_\beta(k)},
\end{equation} 
and the particle-particle,
\begin{equation}
 \overline{\Delta}^{\text{SC}}_{\alpha,\beta} = \frac{1}{\beta V} \int_k \braket{\bar{c}_\alpha(k)\bar{c}_\beta(-k)},
\end{equation}
mean-field parameters. The corresponding linearized self-consistency equations read
\begin{equation}
 \Delta_{\alpha,\beta}^{\text{DW}} \sim \int_k G_{\alpha}(k) \, \delta_{\alpha,\beta} + 2\Delta_{\alpha',\beta'}^{\text{DW}}\mathcal{W}_{\alpha \beta'}^{\alpha' \beta} \int_k G_{\alpha}(k)G_{\beta}(k) \label{MFequParticleHole}
\end{equation}
and
\begin{equation}
 \overline{\Delta}_{\alpha,\beta}^{\text{SC}} \sim  2\mathcal{W}_{\alpha \beta}^{\alpha' \beta'} \overline{\Delta}_{\beta',\alpha'}^{\text{SC}} \int_k G_{\alpha}(k)G_{\beta}(k) \label{MFequParticleParticle}
\end{equation}
for the density wave and superconducting order parameters, respectively. Here $G$ denotes the non-interacting Green's function as given in \equref{ShellGreensFunction} without the momentum constraints.
Again focusing on the limit $\eta\rightarrow 0$, we find 
 \begin{align}
  &\int_k G_{\alpha}(k)G_{\beta}(k) =-\frac{\Lambda_{\parallel}}{(2\pi)^2v_\uparrow} \label{J2Result} \\ 
&\times \begin{cases} \frac{f\left(v_{j_\alpha}\Lambda_{\perp}/T\right)+f\left(v_{j_\beta}\Lambda_{\perp}/T\right)}{x_{j_\alpha}+x_{j_\beta}},  & \sigma_\alpha = -\sigma_\beta,  \\ 
\frac{f\left(v_{j_\alpha}\Lambda_{\perp}/T\right)-f\left(v_{j_\beta}\Lambda_{\perp}/T\right)}{x_{j_\alpha}-x_{j_\beta}},  & \sigma_\alpha = \sigma_\beta \wedge v_{j_\alpha} \neq v_{j_\beta}, \\ 
 \frac{\tanh\left(v_{j_\alpha}\Lambda_{\perp}/T\right)}{x_{j_\alpha}},  & \sigma_\alpha = \sigma_\beta \wedge v_{j_\alpha} = v_{j_\beta}, \end{cases} \nonumber 
 \end{align} 
where
\begin{equation}
 f(x) = \int_0^x \frac{\diff \xi}{\xi} \tanh(\xi/2).\\
\end{equation}

As we are interested in the limit $T \ll v_2\Lambda_\perp$, we only keep the leading ($\log$-divergent) terms in the mean-field equations. Therefore, both the first term in \equref{MFequParticleHole} and the cases with $\sigma_\alpha=\sigma_\beta$ in \equref{J2Result} are subdominant. One then finds 
\begin{subequations}
\begin{align}
\begin{split}
 \Delta_{\alpha,\beta}^{\text{DW}} &\sim L_{j_\alpha,j_\beta} \Bigl[ \delta_{\sigma_\alpha,-}\delta_{\sigma_\beta,+} W_{j_\beta',j_\alpha}^{j_\alpha',j_\beta} \Delta_{(-,j_\alpha')(+,j_\beta')}^{\text{DW}} \\ 
&+ \delta_{\sigma_\alpha,+}\delta_{\sigma_\beta,-} W_{j_\alpha,j_\beta'}^{j_\beta,j_\alpha'} \Delta_{(+,j_\alpha')(-,j_\beta')}^{\text{DW}}  \Bigr]  \label{SimplifiedMFequations1}\end{split}
\end{align}
and
\begin{align}
 \overline{\Delta}_{(-,j_\alpha),(+,j_\beta)}^{\text{SC}}&\sim -2L_{j_\alpha,j_\beta} W_{j_\beta,j_\alpha}^{j_\alpha',j_\beta'} \overline{\Delta}_{(-,j_\alpha')(+,j_\beta')}^{\text{SC}}, \label{SimplifiedMFequations2}
\end{align} \label{SimplifiedMFequations}\end{subequations}
where we have introduced 
\begin{equation}
 L_{j_\alpha,j_\beta} := \frac{\left[\log\left(\frac{v_{1}\Lambda_{\perp}}{T}\right)\right]^{x_{j_\alpha}}+\left[\log\left(\frac{v_1\Lambda_{\perp}}{T}\right)\right]^{x_{j_\beta}}}{x_{j_\alpha}+x_{j_\beta}}. \label{GeneralLogarithm}
\end{equation}
In \equref{SimplifiedMFequations2}, it has been exploited that $\overline{\Delta}^{\text{SC}}_{\alpha,\beta}$ is antisymmetric such that it is sufficient to consider $(\sigma_\alpha,\sigma_\beta)=(-,+)$.
We see that, both for the density wave and for the superconducting channel, solely order parameters with $\sigma_\alpha=-\sigma_\beta$ are relevant. In addition, only the backscattering tensor $W$ enters, whereas forward scattering, $V$, does not play any role at all.  

Next, let us expand the density wave order parameters, 
\begin{equation}
 \Delta_{\alpha,\beta}^{\text{DW}} = \sum_{i,j=0}^{3} c_{i,j} \left(\tau_i\right)_{\sigma_\alpha,\sigma_\beta}\left(\sigma_j\right)_{j_\alpha,j_\beta}, \qquad c_{i,j} \in \mathbbm{R}, \label{ParticleHoleOP}
\end{equation} 
and the anomalous expectation values,
\begin{equation}
 \overline{\Delta}_{(-,j_\alpha),(+,j_\beta)}^{\text{SC}}= \sum_{j=0}^3 \tilde{c}_j\left(\sigma_j\right)_{j_\alpha,j_\beta}, \qquad \tilde{c}_j \in \mathbbm{C}, \label{ParticleParticleOP}
\end{equation}
in Pauli matrices to rewrite \equref{SimplifiedMFequations} more explicitly for the three different scenarios shown in \figref{FermiSurfaceConventions}(b)-(d).

\renewcommand{\arraystretch}{1.9}
\begin{table*}[tb]
\caption{Mean-field equations for the density wave, \equref{ParticleHoleOP}, and superconducting order parameters, \equref{ParticleParticleOP}, in case of disjoint regions in momentum space and identical velocities. The plus (minus) sign in the column of $R^{c2}$ and $\Theta$ means that the corresponding order parameter is symmetric (antisymmetric) under $\pi$-rotation and time-reversal, respectively. The mean-field equations with $j=1$ and $j=2$ are degenerate.}
\label{MFEqus1}
 \begin{tabular}{c|c|c|c|c} \hline \hline
Mean-field equations ($j=1,2$) & Order parameter & $R^{c2}$ & $\Theta$ & Phase \\ \hline 
 $\log\left(\frac{v\Lambda_{\perp}}{T}\right)\begin{pmatrix} g_{00} + g_{33} & 2g_{30} \\ 2g_{30} & g_{00}+g_{33} \end{pmatrix} \begin{pmatrix} c_{j,0} \\ c_{j,3} \end{pmatrix} = \begin{pmatrix} c_{j,0} \\ c_{j,3} \end{pmatrix}$ & $\begin{pmatrix}\tau_1\sigma_0 \\ \tau_1\sigma_3\end{pmatrix}$, $\begin{pmatrix}\tau_2\sigma_0 \\ \tau_2\sigma_3\end{pmatrix}$ & $-$, $+$ & $-$, $-$ & $\begin{matrix} SDW^{11} (g_{30}>0)/ \vspace{-0.6em} \\ SDW^{22} (g_{30}<0) \end{matrix}$   \\

$\log\left(\frac{v\Lambda_{\perp}}{T}\right) \left(g_{00} + 2g_{11} - g_{33}\right)c_{j,1} = c_{j,1}$  &  $\tau_1\sigma_1$, $\tau_2\sigma_1$ & $-, +$ & $-, -$ & $SDW^{12}$ \\ 

$\log\left(\frac{v\Lambda_{\perp}}{T}\right) \left(g_{00} - 2g_{11} - g_{33}\right)c_{j,2} = c_{j,2}$  &  $\tau_1\sigma_2$, $\tau_2\sigma_2$ & $-, +$ & $+, +$ & $CDW^{12}$\\ \hline 

$-2\log\left(\frac{v\Lambda_{\perp}}{T}\right)\begin{pmatrix} g_{00} + 2g_{11} + g_{33} & 2g_{30} \\ 2g_{30} & g_{00} - 2g_{11} + g_{33} \end{pmatrix} \begin{pmatrix} \tilde{c}_0 \\ \tilde{c}_{3} \end{pmatrix} = \begin{pmatrix} \tilde{c}_0 \\ \tilde{c}_{3} \end{pmatrix}$ & $\begin{pmatrix} \sigma_0 \\ \sigma_3 \end{pmatrix}$ & $+$ & $+$ & $SC^{++}$/$SC^{+-}$  \\ 

$2\log\left(\frac{v\Lambda_{\perp}}{T}\right) \left(g_{33}-g_{00}\right)\tilde{c}_j = \tilde{c}_j$  &  $\sigma_1$, $\sigma_2$ & $+, -$ & $+, +$ & $SC^{12}$ \\ \hline \hline
\end{tabular}
\end{table*}

\renewcommand{\arraystretch}{1.9}
\begin{table*}[tb]
\caption{Mean-field equations in case of non-overlapping regions in momentum space and different velocities, $v_1>v_2$.}
\label{MFEqus2}
 \begin{tabular}{c|c|c|c|c} \hline \hline
Mean-field equations ($j=1,2$) & Order parameter & $R^{c2}$ & $\Theta$ & Phase \\ \hline 
 $\log\left(\frac{v_1\Lambda_{\perp}}{T}\right)\left(g_{00}+2g_{30}+g_{33}\right)c_{j,0} = c_{j,0}, \quad c_{j,0}=c_{j,3}$ & $\begin{matrix} \tau_1(\sigma_0+\sigma_1), \vspace{-0.8em} \\ \tau_2(\sigma_0+\sigma_1)\end{matrix}$ & $-, +$ & $-, -$ & $SDW^{11}$   \\

$\log\left(\frac{v_1\Lambda_{\perp}}{T}\right) \frac{v_1}{v_1+v_2} \left(g_{00} + 2g_{11} - g_{33}\right)c_{j,1} = c_{j,1}$  &  $\tau_1\sigma_1$, $\tau_2\sigma_1$ & $-, +$ & $-, -$ & $SDW^{12}$ \\ 

$\log\left(\frac{v_1\Lambda_{\perp}}{T}\right) \frac{v_1}{v_1+v_2} \left(g_{00} - 2g_{11} - g_{33}\right)c_{j,2} = c_{j,2}$  &  $\tau_1\sigma_2$, $\tau_2\sigma_2$ & $-, +$ & $+, +$ & $CDW^{12}$\\ \hline 

$-2\log\left(\frac{v_1\Lambda_{\perp}}{T}\right)(g_{00}+2g_{30}+g_{33})\tilde{c}_0  = \tilde{c}_0, \quad \tilde{c}_0=\tilde{c}_3$ & $ \sigma_0 + \sigma_3 $ & $+$ & $+$ & $SC^{11}$  \\ 

$2\log\left(\frac{v_1\Lambda_{\perp}}{T}\right) \frac{v_1}{v_1+v_2}  \left(g_{33}-g_{00}\right)\tilde{c}_j = \tilde{c}_j$  &  $\sigma_1$, $\sigma_2$ & $+, -$ & $+, +$ & $SC^{12}$ \\ \hline \hline
\end{tabular}
\end{table*}

\subsubsection{Instabilities for identical velocities and disjoint momentum spaces}
By definition, we have $x_j=1$ in this case and hence
\begin{equation}
 L_{j_\alpha,j_\beta} = \log\left(\frac{v\Lambda_{\perp}}{T}\right). \label{TheSimplestL}
\end{equation}
Inserting the coupling matrix (\ref{GeneralgDisjoint}) into \equref{SimplifiedMFequations}, we find the mean-field equations summarized in \tableref{MFEqus1}. 

To discuss the implications of this result, let us first assume that the couplings diverge before the RG flow is cut off due to the finite curvature of the Fermi surface. In regime (I) of \figref{RGflowAndPhaseDiagram}(a), the couplings behave asymptotically as $g_{11} \sim -g_{33} \rightarrow \infty$, whereas $g_{00}$ and $g_{30}$ stay finite. As is easily seen from the mean-field equations, the leading instability is, in this case, characterized by $\overline{\Delta}_{(-,j_\alpha),(+,j_\beta)}^{\text{SC}} \propto \left(\sigma_3\right)_{j_\alpha,j_\beta}$. Thus, the system resides in the $SC^{+-}$-state. Correspondingly, in regime (II), we have $g_{11} \sim g_{33} \rightarrow -\infty$ and hence $\overline{\Delta}_{(-,j_\alpha),(+,j_\beta)}^{\text{SC}} \propto \left(\sigma_0\right)_{j_\alpha,j_\beta}$, i.e.~$SC^{++}$, dominates.
To derive the subleading instabilities, we have investigated the flow of all mean-field equations in \tableref{MFEqus1} according to \equref{RGflowequations1} and analyzed which of the order parameters is dominant before superconductivity eventually wins. Since, at that point, $g_{11}$ and $g_{33}$ are still finite, the result also depends on the value of the non-flowing coupling constants. The associated instabilities, that compete with $SC^{+-}$ and $SC^{++}$, are shown in \figref{RGflowAndPhaseDiagram}(c) and (d) of the main text.

\renewcommand{\arraystretch}{1.9}
\begin{table*}[t]
\caption{Mean-field equations in case of overlapping regions in momentum space. The transformation behavior of the order parameters under $R^{c2}$ and $\Theta$ can be found in \tableref{MFEqus1}.}
\label{MFEqus3}
 \begin{tabular}{c|c} \hline \hline
Mean-field equations ($j=1,2$) & Order parameter \\ \hline 
$\log\left(\frac{v\Lambda_{\perp}}{T}\right)\begin{pmatrix} g_{00} + g_{11} + g_{22} + g_{33} & 2g_{10} & 2g_{30} \\ 2g_{10} & g_{00} + g_{11} - g_{22} - g_{33} & 0 \\ 2g_{30} & 0 & g_{00} - g_{11} - g_{22} + g_{33} \end{pmatrix} \begin{pmatrix} c_{j,0} \\ c_{j,1} \\ c_{j,3} \end{pmatrix} = \begin{pmatrix} c_{j,0} \\ c_{j,1} \\ c_{j,3} \end{pmatrix}$ & $\begin{pmatrix}\tau_1\sigma_0 \\ \tau_1\sigma_1 \\ \tau_1\sigma_3\end{pmatrix}$, $\begin{pmatrix}\tau_2\sigma_0 \\ \tau_2\sigma_1 \\ \tau_2\sigma_3\end{pmatrix}$  \\
$\log\left(\frac{v\Lambda_{\perp}}{T}\right)  \left(g_{00} -g_{11} + g_{22} - g_{33}\right)c_j = c_j$ &  $\tau_1\sigma_2$, $\tau_2\sigma_2$ \\ \hline

$-2\log\left(\frac{v\Lambda_{\perp}}{T}\right) \begin{pmatrix} g_{00}+g_{11}-g_{22}+g_{33} & 2g_{10} & 2g_{30} \\ 2g_{10} & g_{00}+g_{11}+g_{22}-g_{33} & 0 \\ 2g_{30} & 0 & g_{00}-g_{11}+g_{22}+g_{33} \end{pmatrix} \begin{pmatrix} \tilde{c}_0 \\ \tilde{c}_1 \\ \tilde{c}_{3} \end{pmatrix} = \begin{pmatrix} \tilde{c}_0 \\ \tilde{c}_1 \\ \tilde{c}_{3} \end{pmatrix}$ & $\begin{pmatrix} \sigma_0 \\ \sigma_1 \\ \sigma_3 \end{pmatrix}$ \\ \hline

$2\log\left(\frac{v\Lambda_{\perp}}{T}\right) \left(-g_{00}+g_{11}+g_{22}+g_{33}\right)\tilde{c}_2 = \tilde{c}_2$  &  $\sigma_2$ \\ \hline
\hline
\end{tabular}
\end{table*}

\subsubsection{Different Fermi velocities}
If $v_1 > v_2$, we only take the leading logarithm in \equref{GeneralLogarithm}, i.e.~
\begin{equation}
 L_{j_\alpha,j_\beta} \sim \log\left(\frac{v_1\Lambda_{\perp}}{T}\right) \left[\delta_{j_\alpha,1}\delta_{j_\beta,1} +  \frac{1-\delta_{j_\alpha,j_\beta}}{1+x_2}\right]
\end{equation} 
yielding the mean-field equations presented in \tableref{MFEqus2}. Note that, as a consequence of the asymmetry between the Fermions from the inner and outer Fermi surfaces, the instabilities $SDW^{11}$ and $SC^{11}$ dominate over $SDW^{22}$ and $SC^{22}$, respectively. Using the RG flow in \equref{RGflowequations2}, one can easily determine the phase diagram. Remarkably, it turns out that superconductivity will still be the leading instability if the RG is not cut off before the backscattering coupling constants diverge. The difference, compared to the situation with identical velocities, is that the order parameter of the resulting superconductor ($SC^{11}$) is 
\begin{equation}
 \overline{\Delta}_{(-,j_\alpha),(+,j_\beta)}^{\text{SC}} \propto \delta_{j_\alpha,1}\delta_{j_\beta,1}, \label{DifferentVSCInst}
\end{equation} 
both for the conventional and for the unconventional pairing scenario. We emphasize that the resulting mean-field theories in the associated blue and red part of the flow in \figref{AdditionalRGFlows}(a) are not identical as the coupling constants are different. In \secref{InvariantsDifferent}, we show that the superconductors even differ in their topology.

\renewcommand{\arraystretch}{1.4}
\begin{table}[t]
\caption{Instabilities for weak spin-orbit coupling.}
\label{ResultingPhases}
 \begin{tabular} {c|c|c} \hline \hline
Fixed point & Divergence & Order parameter  \\ \hline
(I)  & $g_{11} \sim g_{22} \sim g_{33} \rightarrow \infty$ & $\overline{\Delta}_{(-,\cdot),(+,\cdot)}^{\text{SC}}=\sigma_2$ \\ 
(II)  & $-g_{11} \sim g_{22} \sim g_{33} \rightarrow -\infty$ & $\overline{\Delta}_{(-,\cdot),(+,\cdot)}^{\text{SC}}=\sigma_3$ \\ 
(III)  & $-g_{11} \sim -g_{22} \sim g_{33} \rightarrow \infty$ & $\overline{\Delta}_{(-,\cdot),(+,\cdot)}^{\text{SC}}=\sigma_1$ \\ 
(IV)  & $g_{11} \sim -g_{22} \sim g_{33} \rightarrow -\infty$ & $\overline{\Delta}_{(-,\cdot),(+,\cdot)}^{\text{SC}}=\sigma_0$ \\ \hline \hline
 \end{tabular}
\end{table}

\subsubsection{Quasi-degenerate Fermi surfaces}
Since $v_1=v_2$ in the limit of very weak spin-orbit splitting, $L_{j_\alpha,j_\beta}$ is again given by \equref{TheSimplestL}. Using the reduced coupling matrix (\ref{GeneralgOverlappingRed}), we find the mean-field equations of \tableref{MFEqus3}, where primes have been neglected for notational simplicity.

As the Fermi surfaces are quasi-degenerate, also superconductors with off-diagonal order parameters, $\overline{\Delta}_{(-,j_\alpha),(+,j_\beta)}^{\text{SC}} \neq 0$ for $j_\alpha\neq j_\beta$, are possible.   
Recall from \figref{AdditionalRGFlows}(b), (c) and from \equref{SolutionAtStableFPs} that there are four stable fixed points at which the couplings diverge. From the mean-field equations, the leading instability associated with the divergences at the four fixed points is readily found and summarized in \tableref{ResultingPhases}. Interestingly, as long as curvature corrections of the Fermi surface are negligible, superconductivity generically wins even in the present case with the largest number of independent coupling constants.

\subsection{Detailed calculation of the invariants}
\label{Invariants}
Since, in none of the superconducting phases derived above, time-reversal symmetry $\Theta$ is spontaneously broken by the strongly nested parts of the Fermi surface, it is reasonable to assume that this holds for the entire Fermi surface, if $m_l/m_h$ is sufficiently small. 
Similarly, we only know that the superconducting order parameter is finite in the nested parts of the Fermi surfaces. Since the four equivalent strongly nested subspaces have been treated independently, our analysis does not tell whether the point symmetries relating these subspaces are spontaneously broken and whether the order parameter changes sign along the Fermi surface. This crucially depends on the non-singular interaction channels between the nested subspaces. As the values of the corresponding coupling constants are \textit{a priori} unknown, let us assume that the order parameter is finite on the entire Fermi surface which is motivated by the experimental analysis\cite{Richter} of the superconducting gap of the system.

As $\Theta^2=-\mathbbm{1}$, the superconductor belongs to class DIII and is characterized by a $\mathbbm{Z}_2$ topological invariant in two spatial dimensions.
Here we present the calculation of the invariants in much more detail and for all three regimes of the system.

\subsubsection{Identical velocities}
Again, let us start with the simplest situation of disjoint support in momentum space and identical velocities. As shown above, the resulting superconductors are characterized by
\begin{equation}
 \overline{\Delta}_{(-,j_\alpha),(+,j_\beta)}^{\text{SC}}=-\overline{\Delta}_{(+,j_\beta),(-,j_\alpha)}^{\text{SC}}= \tilde{c} \left(\sigma_j\right)_{j_\alpha,j_\beta}, \label{SCsMeanField}
\end{equation} 
$\tilde{c} \in \mathbbm{C}$, with $j=0$ and $j=3$ in the conventional and unconventional pairing scenario, respectively. Treating the interaction (\ref{GeneralProjectedInteraction}) at mean-field level and inserting \equref{SCsMeanField}, one finds
\begin{align}
 S^{\text{MF}}_{\text{int}} &= \int_{k} \overline{\Delta}_{\alpha',\beta'}^{\text{SC}} \mathcal{W}^{\alpha'\beta'}_{\alpha\beta} c_\alpha(k)c_\beta(-k) + \text{G.c.} \\
&=  \int_{k} c_{(-,j_\alpha)}(-k) m_{j_\alpha,j_\beta} c_{(+,j_\beta)}(k) + \text{G.c.}, \label{GeneralMFmForm}
\end{align} 
where the mean-field parameter
\begin{equation}
 m_{j_\alpha,j_\beta}:= -4\tilde{c} \left(\sigma_j\right)_{j_\alpha',j_\beta'} W^{j_\alpha',j_\beta'}_{j_\beta,j_\alpha} \label{ExplicitFormOfMFParameters}
\end{equation} 
has been introduced. In the present case, with $g$ as given in \equref{GeneralgDisjoint}, we find
\begin{equation}
 m = -4\tilde{c} \left(\gamma_0\sigma_0+\gamma_3\sigma_z\right), \label{mStructure1}
\end{equation} 
where
\begin{subequations}
\begin{align}
 \gamma_0 &= \begin{cases} g_{00} + 2g_{11} + g_{33}, & j=0, \\ 2g_{30}, & j=3,  \end{cases} \\
 \gamma_3 &= \begin{cases} 2g_{30}, & j=0, \\ g_{00} - 2g_{11} + g_{33}, & j=3.  \end{cases}
\end{align}\label{GammaDefinition1}\end{subequations}
Before calculating the $\mathbbm{Z}_2$-invariant, it is instructive to first investigate the excitation spectrum of the effective one-dimensional system. For this purpose, we introduce Nambu spinors
\begin{equation}
 \Phi(k) = \begin{pmatrix} c_{(+,1)}(k) \\ c_{(+,2)}(k) \\ \bar{c}_{(-,1)}(-k) \\ \bar{c}_{(-,2)}(-k) \end{pmatrix}, \quad \overline{\Phi}(k) = \begin{pmatrix} \bar{c}_{(+,1)}(k) \\ \bar{c}_{(+,2)}(k) \\ c_{(-,1)}(-k) \\ c_{(-,2)}(-k) \end{pmatrix}^T \hspace{-0.7em},
\end{equation} 
to write the mean-field action in quadratic form,
\begin{equation}
 S_0 + S_{\text{MF}} = \int_k \overline{\Phi}_i(k) \left(-i\omega_n\delta_{i,j} + \mathcal{H}_{i,j}(k) \right) \Phi_j(k),
\end{equation} 
where
\begin{equation}
 \mathcal{H}(k) = \begin{pmatrix} \begin{matrix} vk_\perp & 0 \\ 0 & vk_\perp \end{matrix} & m^\dagger \\ m & \begin{matrix} -vk_\perp & 0 \\ 0 & -vk_\perp \end{matrix} \end{pmatrix}. \label{BdHHamiltonianOf1DSys}
\end{equation} 
Diagonalizing $\mathcal{H}(k)$ readily yields the four bands
\begin{equation}
 E(k) = \pm\sqrt{(vk)^2+16|\tilde{c}|^2\left( \gamma_0\pm \gamma_3\right)^2} \label{ExcitationSpectrum}
\end{equation} 
characterizing excitations in the superconducting phase. Obviously, the gap closes when
\begin{equation}
 |\gamma_0| = |\gamma_3|, \label{GapClosing1}
\end{equation} 
i.e., if there are topologically distinct phases, they have to be separated by a manifold where \equref{GapClosing1} holds.

To calculate the invariant, we have to relate the effective one-dimensional theory to the full mean-field Hamiltonian,
\begin{align}
\begin{split}
 H = \sum_{\vec{k}}& \Psi^\dagger_\alpha(\vec{k}) H_{\alpha,\beta}(\vec{k}) \Psi_\beta(\vec{k}) \\ 
&+ \frac{1}{2} \sum_{\vec{k}} \left[\Psi_\alpha(-\vec{k})\Delta^\dagger_{\alpha,\beta}(\vec{k})\Psi_\beta(\vec{k}) + \text{H.c.} \right], \label{MFHamiltonian1}\end{split}
\end{align}  
defined on the entire two-dimensional Brillouin zone. 
Suppose that the free Hamiltonian $H$ has been diagonalized by applying the transformation (\ref{IntroductionOfcs}). Here we use the convention that the eigenfunctions $\phi_\alpha$ are sorted for every $\vec{k}$ such that the energy increases with $\alpha$. Furthermore, the phases are fixed by demanding that \equsref{PhaseConvention1}{PhaseConvention2} hold.
Then from \equsref{GeneralMFmForm}{mStructure1}, we know that the pairing term, i.e.~the second line in \equref{MFHamiltonian1}, must have the form
\begin{equation}
 H_{\text{pair}} = \sum_{j=1,2}\sum_{\vec{k}\in S_j} f_j(-\vec{k})m_{j,j}f_j(\vec{k}) + \text{H.c.}  + \dots \,. \label{LowenergyPairing} 
\end{equation} 
Here
\begin{equation}
S_j=\left\{\vec{k},\,|\vec{e}_\parallel(\vec{k}-\vec{k}_j)|<\Lambda_\parallel \wedge |\vec{e}_\perp(\vec{k}-\vec{k}_j)|<\Lambda_\perp \right\}        
\end{equation} 
denotes the strongly nested domain in the vicinity of $\vec{k}_j$ (see \figref{FermiSurfaceConventions}(a)) and the ellipsis stands for the pairing terms in the remainder of the Brillouin zone. Using \equref{IntroductionOfcs} to rewrite the $f$-operators in terms of the $\Psi$-fields, we find
\begin{equation}
 \Delta_{\alpha,\beta}^\dagger(\vec{k}) = 2m_{j,j}\left(\phi_j^*(-\vec{k})\right)_\alpha \left(\phi^*_j(\vec{k})\right)_\beta, \quad \forall\,\vec{k}\in S_j. \label{ExpressionForThePairingField}
\end{equation} 
By construction, $H$ is time-reversal invariant, which means that
\begin{equation}
 \theta_H H(\vec{k}) \theta_H^{-1} = H(-\vec{k}), \quad \theta_H = e^{i\varphi}i\sigma_y\mathcal{K} \label{GeneralRepOfTR}
\end{equation} 
with $\mathcal{K}$ denoting complex conjugation and arbitrary $\varphi\in\mathbbm{R}$. It is straightforward to show that time-reversal invariance of the pairing term is equivalent to
\begin{equation}
 e^{2i\varphi}\sigma_y\Delta^\dagger(\vec{k})\sigma_y = -\Delta(\vec{k}). \label{TimeReversalInvarianceOfPairing}
\end{equation} 
Using the phase conventions (\ref{PhaseConvention1}) and (\ref{PhaseConvention2}) and writing $\tilde{c}=|\tilde{c}|e^{i\rho}$, $\rho\in\mathbbm{R}$, one finds that \equref{TimeReversalInvarianceOfPairing} is satisfied if 
\begin{equation}
 \rho+\varphi = \pm \frac{\pi}{2}. \label{GeneralSolutionOfTRSCondition}
\end{equation} 
Now we are prepared to calculate the $\mathbbm{Z}_2$-invariant. According to \refcite{Zhang}, one simply has to evaluate the matrix elements
\begin{equation}
 \delta_j := \braket{\phi_j(\vec{q}_j)|\theta_H\mathcal{K}\Delta^\dagger(\vec{q}_j)|\phi_j(\vec{q}_j)} \in\mathbbm{R}, \label{TheMatrixElements}
\end{equation}  
where $\vec{q}_j$ is an arbitrary point on the $j$th Fermi surface. As long as the gap of the superconducting system does not close, the sign of $\delta_j$ is constant on the entire Fermi surface\cite{Zhang} and we are free to choose $\vec{q}_j=\vec{k}_j$. From \equref{ExpressionForThePairingField}, one then finds
\begin{align}
 \delta_j &= 2ie^{i\varphi} m_{j,j}\left(\phi_j^\dagger(\vec{k}_j)\sigma_y\phi_j^*(-\vec{k}_j)\right) \left(\phi^\dagger_j(\vec{k}_j)\phi_j(\vec{k}_j)\right) \nonumber \\
&= 2ie^{i\varphi} m_{j,j}\phi_j^\dagger(\vec{k}_j)\sigma_y\sigma_x\phi_j(-\vec{k}_j) \nonumber \\
&= -2ie^{i\varphi} m_{j,j}\phi_j^\dagger(\vec{k}_j)\phi_j(\vec{k}_j) \nonumber \\
&= -2ie^{i\varphi} m_{j,j}, \nonumber \\
&= \mp 8 |\tilde{c}| \begin{cases} \gamma_0 + \gamma_3, & j=1, \\ \gamma_0 - \gamma_3, & j=2, \end{cases} \label{ResultForDeltaIdVel}
\end{align}
again exploiting \equsref{PhaseConvention1}{PhaseConvention2} as well as, in the last line, \equsref{mStructure1}{GeneralSolutionOfTRSCondition}.
As both Fermi surfaces enclose one time-reversal invariant point, the invariant is given by\cite{Zhang}
\begin{align}
 N &= \prod_j \sign(\delta_j) = \begin{cases} +1 \text{ (trivial)}, & |\gamma_0|>|\gamma_3|, \\ -1 \text{ (nontrivial)}, & |\gamma_0|<|\gamma_3|. \end{cases} \label{TopTrivPhaseBoundary}
\end{align}
First, note that the phase boundary $|\gamma_0|=|\gamma_3|$ is in accordance with the analysis of the excitation spectrum (\ref{ExcitationSpectrum}) of the one-dimensional description. Secondly, let us rewrite the condition for a topologically nontrivial phase considering the conventional and the unconventional superconductor separately. In the latter case, the system is topological if
\begin{equation}
 |2g_{30}|<|g_{00}-2g_{11}+g_{33}|
\end{equation} 
and the RG flow (see region (I) in \figref{RGflowAndPhaseDiagram}(a)) leads to the asymptotic behavior $g_{11}\sim -g_{33} \rightarrow \infty$, whereas $g_{00}$ and $g_{30}$ do not flow at all. Consequently, the unconventional superconductor $SC^{+-}$ is topologically nontrivial.

In case of conventional pairing, though, the condition for $N=-1$ reads
\begin{equation}
  |g_{00}+2g_{11}+g_{33}|<|2g_{30}|
\end{equation} 
and the RG flow (regime (II) in \figref{RGflowAndPhaseDiagram}(a)) behaves asymptotically as $g_{11}\sim g_{33} \rightarrow -\infty$. Therefore, the conventional $SC^{++}$-phase is trivial.

\subsubsection{Generalization to different velocities}
\label{InvariantsDifferent}
In this case, only one superconducting order parameter, as given in \equref{DifferentVSCInst}, is possible. From \equref{ExplicitFormOfMFParameters}, we can immediately conclude that
\begin{equation}
 m_{j_\alpha,j_\beta}=-4\tilde{c} \left(\sigma_0+\sigma_3\right)_{j_\alpha',j_\beta'} W^{j_\alpha',j_\beta'}_{j_\beta,j_\alpha} = -8\tilde{c} \, W^{1,1}_{j_\beta,j_\alpha}
\end{equation} 
and, hence, $\gamma_0 = g_{00}+2g_{11}+g_{33}+2g_{30}$, $\gamma_3 = g_{00}-2g_{11}+g_{33}+2g_{30}$ using the parameterization (\ref{mStructure1}). The condition $|\gamma_0|<|\gamma_3|$ for having a topologically nontrivial superconductor then becomes
\begin{equation}
 |g_{00}+2g_{11}+g_{33}+2g_{30}| < |g_{00}-2g_{11}+g_{33}+2g_{30}|. \label{DifferentFVsTopCond}
\end{equation} 
Recall from the analysis of the flow equations (\ref{RGflowequations2}) that $g_{00}$, $g_{33}$ and $g_{30}$ can only diverge to $-\infty$, whereas $g_{11}\rightarrow \infty$ in case of unconventional pairing (regime (I) in \figref{AdditionalRGFlows}(a)) and $g_{11}\rightarrow -\infty$ for the conventional superconductor (regime (II)).
Hence, we can easily see from \equref{DifferentFVsTopCond}, that, exactly as above, the conventional superconductor is trivial and the unconventionally paired state is topological.

\renewcommand{\arraystretch}{1.6}
\begin{table}[b]
\caption{Mean-field parameters as defined in \equref{GeneralMFmForm} for overlapping domains in momentum space.}
\label{MeanFieldOverlapping}
\begin{tabular}{c|c} \hline\hline
Fixed point & $m=-4\tilde{c}(\gamma_0\sigma_0 + \vec{\gamma}\cdot\vec{\sigma})$  \\ \hline 
(I) & $\gamma_2=g_{00}-g_{11}-g_{22}-g_{33}$, $\gamma_0=\gamma_1=\gamma_3=0$ \\ 
(II) & $\begin{matrix} \gamma_0=2g_{30}, \gamma_3=g_{00}-g_{11}+g_{22}+g_{33}, \vspace{-0.6em} \\ \gamma_1=\gamma_2=0 \end{matrix}$ \\ 
(III) & $\begin{matrix} \gamma_0=2g_{10}, \gamma_1=g_{00}+g_{11}+g_{22}-g_{33}, \vspace{-0.6em} \\ \gamma_2=\gamma_3=0 \end{matrix}$ \\ 
(IV) & $\begin{matrix} \gamma_0=g_{00}+g_{11}-g_{22}+g_{33}, \gamma_1=2g_{10}, \vspace{-0.6em} \\ \gamma_2=\gamma_3=0 \end{matrix}$ \\ \hline\hline
\end{tabular}
\end{table}

\subsubsection{Quasi-degenerate Fermi surfaces}
Finally, we also discuss the situation of very weak spin-orbit coupling ($0<\eta\ll v_j\Lambda_\perp$), where four distinct superconductors are possible as summarized in \tableref{ResultingPhases}. From \equref{ExplicitFormOfMFParameters} and recalling the reduced backscattering coupling matrix (\ref{GeneralgOverlappingRed}), we find the mean-field parameters in \tableref{MeanFieldOverlapping} associated with the four stable fixed points in \figref{AdditionalRGFlows}(b) and (c). 

Calculating the excitation spectrum of the corresponding one-dimensional BdG Hamiltonian,
\begin{equation}
 \mathcal{H}(k) = \begin{pmatrix} \begin{matrix} vk_\perp+\eta & 0 \\ 0 & vk_\perp-\eta \end{matrix} & -4\tilde{c}^*(\gamma_0\sigma_0 + \vec{\gamma}\cdot\vec{\sigma}) \\ -4\tilde{c}(\gamma_0\sigma_0 + \vec{\gamma}\cdot\vec{\sigma}) & \begin{matrix} -vk_\perp-\eta & 0 \\ 0 & -vk_\perp+\eta \end{matrix} \end{pmatrix},
\end{equation} 
one finds that the off-diagonal components, $\gamma_1$, $\gamma_2$, of $m$ do not open up a gap for $|\tilde{c}\gamma_s|\ll \eta$. Consequently, the superconductors associated with the fixed points (I) and (III) are (asymptotically) gapless. In the former case $m$ is exactly off-diagonal, whereas in the latter, $m$ becomes off-diagonal as $\gamma_0/\gamma_1\rightarrow 0$ when the couplings diverge according to $-g_{11} \sim -g_{22} \sim g_{33} \rightarrow \infty$ at the fixed point (III). Consequently, the $\mathbbm{Z}_2$-invariant is not defined in these two cases. The superconductors corresponding to the fixed points (II) and (IV), however, are indeed gapped for infinitesimal $\tilde{c}\gamma_s$ as the associated matrices $m$ have finite diagonal components.

To investigate the topological properties of these states, note that, for any finite $\eta$, we can still sort the eigenfunctions by energy and then apply the phase conventions in \equsref{PhaseConvention1}{PhaseConvention2}. Similarly to the above analysis, one can then rewrite the matrix elements
\begin{equation}
 \braket{\phi_j(\vec{k}_1)|\theta_H\mathcal{K}\Delta^\dagger(\vec{k}_1)|\phi_{j'}(\vec{k}_1)} = \mp 8 |\tilde{c}|\left(\gamma_0\sigma_0 + \vec{\gamma}\cdot\vec{\sigma}\right), \label{generalMatrixElements}
\end{equation} 
$j=1,2$, in terms of the mean-field parameters $\gamma_s$. As in \equref{ResultForDeltaIdVel}, the two possible signs correspond to the two possible choices of $\varphi$ in the representation (\ref{GeneralRepOfTR}) of time-reversal.

For the procedure of \refcite{Zhang} for calculating the topological invariant to work, it is essential that the matrix elements in \equref{generalMatrixElements} between different Fermi surfaces can be neglected. For the superconductors (II) and (IV) this is indeed valid since, in the first case, $m$ is exactly diagonal, and, in the latter, $m$ becomes asymptotically diagonal when the couplings diverge. From \equref{generalMatrixElements}, it is readily seen that the condition for having a topologically nontrivial superconductor is again given by $|\gamma_0|<|\gamma_3|$. Using the mean-field parameters defined in \tableref{MeanFieldOverlapping} and recalling \equref{SolutionAtStableFPs}, one finds that the superconductors associated with the fixed points (II) and (IV) are topological and trivial, respectively.

Consequently, even in the scenario of quasi-degenerate Fermi surfaces, only two fully gapped superconductors are possible: The unconventional superconductor (red region in \figref{AdditionalRGFlows}(b), flowing to (II) in \figref{AdditionalRGFlows}(c)) has a nontrivial $\mathbbm{Z}_2$-invariant, whereas the conventionally paired state (blue region in \figref{AdditionalRGFlows}(c)) is trivial.

\subsection{Spatial structure of the density waves}
In the following, we present more details about how the density wave profiles in \figref{IllustrationOfDWPhases} have been derived.
The charge ($s=0$) and spin ($s=1,2,3$) expectation value is given by
\begin{subequations}
\begin{align}
 \mathcal{S}_s(\vec{x},\tau) &= \int_q \hat{\mathcal{S}}_s(q) e^{i(\vec{q}\cdot\vec{x}-\omega_n\tau)}, \\
 \hat{\mathcal{S}}_s(q) &= \int_k\braket{\overline{\Psi}(k)\sigma_s \Psi(k+q)},
\end{align}\end{subequations}
where $\tau$ denotes (imaginary) time, $\Psi$ are the four component fields as in \equref{nondiagonalizedquadraticaction} and $\sigma_s$ act in spin-space. Applying the transformation (\ref{IntroductionOfcs}), one can write
\begin{equation}
 \hat{\mathcal{S}}_s(q) = \int_k \phi_\alpha^\dagger(\vec{k})\sigma_s\phi_\beta(\vec{k}+\vec{q})  \braket{\bar{f}_\alpha(k)f_\beta(k+q)}. \label{SInQSpace}
\end{equation} 
Let us assume that the mass anisotropy is sufficiently large such that the contribution to $\mathcal{S}_s$ of the strongly curved segments of the Fermi surface is negligible. Recall from \figref{SpectrumAndWavefunctions}(b) and (c) that the orbital part of the wave functions is strongly polarized to either $xz$ or $yz$ in the remaining nearly straight segments of the Fermi surface. For this reason, the orbital momentum $L_j$ cannot contribute to the magnetization of the sample as the matrix elements $\braket{xz|L_j|xz}$ and $\braket{yz|L_j|yz}$ vanish for all components $j=1,2,3$.

Let us first focus on the spin-density wave $SDW^{12}$, which is characterized by $\Delta^{\text{DW}}=c \, \tau_1\sigma_1$ or $\Delta^{\text{DW}}= c \, \tau_2\sigma_1$, $c\in\mathbbm{R}$. As explained in the main text, we focus, for concreteness, on the latter choice as it does not break $\pi$-rotation symmetry. However, using the procedure presented in this section, it is straightforward to derive the spatial texture when some of the point symmetries are broken spontaneously.

The contribution of the red parts in \figref{FermiSurfaceConventions}(a) is then readily found from \equref{SInQSpace},
\begin{align}
\begin{split}
 \mathcal{S}_s(\vec{x},\tau) =& i c\, e^{i \vec{Q}^{(1)}_{12}\cdot\vec{x}} \Bigl[ \phi_1^\dagger(-\vec{k}_1)\sigma_s\phi_2(\vec{k}_2) \\
&+ \phi_2^\dagger(-\vec{k}_2)\sigma_s\phi_1(\vec{k}_1) \Bigr]  + \text{c.c.} + \dots,\end{split} \label{SsCalcPre}
\end{align} 
where we have introduced the corresponding nesting vector $\vec{Q}^{(1)}_{12}=\vec{k}_1+\vec{k}_2$ with $\vec{k}_j$ parameterizing the centers of the nested subspaces (see \figref{RGflowAndPhaseDiagram}(a)). Using the phase conventions (\ref{PhaseConvention1}) and (\ref{PhaseConvention2}), one can show that $\mathcal{S}_0=0$, as required since the order parameter of $SDW^{12}$ is odd under time-reversal, and simplify \equref{SsCalcPre} to
\begin{align}
 \begin{split}
\vec{\mathcal{S}}(\vec{x},\tau) = 2i c\, e^{i \vec{Q}^{(1)}_{12}\cdot\vec{x}}  &\phi_1^\dagger(\vec{k}_1)\left(-\sigma_2, \sigma_1, i\sigma_0\right)^T \phi_2(\vec{k}_2) \\
  &+ \text{c.c.} + \dots\, . \end{split}
\end{align} 
From \figref{SpectrumAndWavefunctions}(b) and (c), it is easily seen that the wave functions are strongly spin-polarized such that $\sigma_2\phi_1(\vec{k}_1)\approx -\phi_1(\vec{k}_1)$ and $\sigma_2\phi_2(\vec{k}_2)\approx \phi_2(\vec{k}_2)$. Within this approximation and, again, using \equref{PhaseConvention2}, we find
\begin{equation}
 \vec{\mathcal{S}}(\vec{x},\tau) \propto \left(0,\sin\left(\vec{Q}^{(1)}_{12}\cdot\vec{x}\right),0\right)^T.
\end{equation} 
Assuming that none of the additional point symmetries are broken spontaneously, we can directly infer the contributions of the other three nested subspaces (blue regions in \figref{FermiSurfaceConventions}(a)). Demanding that $\vec{S}$ transform as a pseudo vector under reflection at the $xz$-plane and under $\pi/2$-rotation around the $z$-axis, one finds
\begin{equation}
 \vec{\mathcal{S}}(\vec{x},\tau) \propto \begin{pmatrix} \sin\left(\vec{Q}^{(3)}_{12}\cdot\vec{x}\right) + \sin\left(\vec{Q}^{(4)}_{12}\cdot\vec{x}\right) \\ \sin\left(\vec{Q}^{(1)}_{12}\cdot\vec{x}\right) + \sin\left(\vec{Q}^{(2)}_{12}\cdot\vec{x}\right) \\ 0 \end{pmatrix}, \label{SDW12Texture}
\end{equation} 
where the reflected and rotated nesting vectors
\begin{subequations}
\begin{align}
\vec{Q}_{12}^{(2)} &= \left(Q_{12,x}^{(1)},-Q_{12,y}^{(1)}\right)^T, \\ 
\vec{Q}_{12}^{(3)} &= \left(Q_{12,y}^{(1)},-Q_{12,x}^{(1)}\right)^T, \\
\vec{Q}_{12}^{(4)} &= \left(-Q_{12,y}^{(1)},-Q_{12,x}^{(1)}\right)^T,
\end{align} \label{DefinitionOfQs}\end{subequations}
have been defined. \equref{SDW12Texture} has been plotted in \figref{IllustrationOfDWPhases}(b) for a specific choice of $\vec{Q}_{12}^{(1)}$.

In the same way, one obtains
\begin{equation}
 \mathcal{S}_0(\vec{x},\tau) \propto \sum_{j=1}^4\cos\left(\vec{Q}_{12}^{(j)}\cdot\vec{x}\right)
\end{equation} 
for the $CDW^{12}$ and ($j=1,2$)
\begin{widetext}
\begin{equation}
 \vec{\mathcal{S}}(\vec{x},\tau) \propto \begin{pmatrix} \sin\left(\vec{Q}^{(1)}_{jj}\cdot\vec{x}\right)-\sin\left(\vec{Q}^{(2)}_{jj}\cdot\vec{x}\right) \\ \sin\left(\vec{Q}^{(4)}_{jj}\cdot\vec{x}\right)-\sin\left(\vec{Q}^{(3)}_{jj}\cdot\vec{x}\right) \\ (-1)^{j+1} \sum\limits_{s=1,2} \left[ \cos\left(\vec{Q}^{(2s-1)}_{jj}\cdot\vec{x}\right)-\cos\left(\vec{Q}^{(2s)}_{jj}\cdot\vec{x}\right) \right] \end{pmatrix}
\end{equation} 
in case of the $SDW^{jj}$-phase, where $\vec{Q}^{(1)}_{jj}=2\vec{k}_j$ and $\vec{Q}^{(p)}_{jj}$, $p=2,3,4$, as defined similarly to \equref{DefinitionOfQs}.
\end{widetext}
\end{document}